\documentclass[sigconf]{acmart}

\usepackage{booktabs} 

\usepackage[english]{babel}
\usepackage{moresize}
\usepackage{amsmath}
\usepackage{algorithmic}
\usepackage{balance}
\usepackage{comment}
\usepackage{paralist}
\usepackage{bm}
\usepackage{pgfplots}
\usetikzlibrary{pgfplots.dateplot}

\usepackage{flushend}
\usepackage[english]{babel}
\usepackage[latin1]{inputenc}
\usepackage{mathrsfs}
\usepackage{graphicx}

\usepackage{amssymb}
\usepackage{amsfonts}
\usepackage{url}
\usepackage{longtable}
\usepackage{rotating}
\usepackage{multirow}
\usepackage{mathrsfs}
\usepackage{subfigure}
\usepackage{enumitem}
\usepackage[linesnumbered,algoruled,boxed,lined]{algorithm2e}
\usepackage{adjustbox}
\usepackage{hyperref}

\usepackage{pgfplots}
\usetikzlibrary{pgfplots.dateplot}

\usepackage{filecontents}
\definecolor{tblue}{RGB}{31,119,180}
\definecolor{torange}{RGB}{255,127,14}
\definecolor{tgreen}{RGB}{44,160,44}
\definecolor{tred}{RGB}{214,39,40}
\definecolor{tpurple}{RGB}{148,103,189}

\newcommand{\hide}[1]{} 

\newcommand{\ie}{\textit{i}.\textit{e}.}
\newcommand{\eg}{\textit{e}.\textit{g}.} 
\newcommand{\wrt}{\textit{w}.\textit{r}.\textit{t}}

\setcopyright{acmcopyright}

\def\model{HCCF}

\begin{document}
\fancyhead{}


\title{Hypergraph Contrastive Collaborative Filtering}

\author{Lianghao Xia}
\affiliation{University of Hong Kong}
\email{aka\_xia@foxmail.com}

\author{Chao Huang}
\authornote{Chao Huang is the corresponding author.}
\affiliation{University of Hong Kong}
\email{chaohuang75@gmail.com}

\author{Yong Xu}
\affiliation{South China University of Technology}
\email{yxu@scut.edu.cn}

\author{Jiashu Zhao}
\affiliation{Wilfrid Laurier University}
\email{jzhao@wlu.ca}

\author{Dawei Yin}
\affiliation{Baidu Inc.}
\email{yindawei@acm.org}

\author{Jimmy Xiangji Huang}
\affiliation{York University}
\email{jhuang@yorku.ca}

\begin{abstract}
Collaborative Filtering (CF) has emerged as fundamental paradigms for parameterizing users and items into latent representation space, with their correlative patterns from interaction data. Among various CF techniques, the development of GNN-based recommender systems, \eg, PinSage and LightGCN, has offered the state-of-the-art performance. However, two key challenges have not been well explored in existing solutions: i) The over-smoothing effect with deeper graph-based CF architecture, may cause the indistinguishable user representations and degradation of recommendation results. ii) The supervision signals (\ie, user-item interactions) are usually scarce and skewed distributed in reality, which limits the representation power of CF paradigms. To tackle these challenges, we propose a new self-supervised recommendation framework \underline{H}ypergraph \underline{C}ontrastive \underline{C}ollaborative \underline{F}iltering (\model) to jointly capture local and global collaborative relations with a hypergraph-enhanced cross-view contrastive learning architecture. In particular, the designed hypergraph structure learning enhances the discrimination ability of GNN-based CF paradigm, so as to comprehensively capture the complex high-order dependencies among users. Additionally, our \model\ model effectively integrates the hypergraph structure encoding with self-supervised learning to reinforce the representation quality of recommender systems, based on the hypergraph-enhanced self-discrimination. Extensive experiments on three benchmark datasets demonstrate the superiority of our model over various state-of-the-art recommendation methods, and the robustness against sparse user interaction data. Our model implementation codes are available at https://github.com/akaxlh/HCCF.
\end{abstract}


%
%
%


\keywords{Recommendation; Collaborative Filtering; Self-Supervised Learning}


\copyrightyear{2022}
\acmYear{2022}
\setcopyright{acmlicensed}\acmConference[SIGIR'22]{Proceedings of the 45th International ACM SIGIR Conference on Research and Development in Information Retrieval}{July 11--15, 2022}{Madrid, Spain}
\acmBooktitle{Proceedings of the 45th International ACM SIGIR Conference on Research and Development in Information Retrieval (SIGIR'22), July 11--15, 2022, Madrid, Spain}
\acmPrice{15.00}
\acmDOI{10.1145/3477495.3532058}
\acmISBN{978-1-4503-8732-3/22/07}

\ccsdesc[500]{Information systems~Recommender systems}

\maketitle

\section{Introduction}
\label{sec:intro}

Personalized recommender systems have been widely utilized to help users discover items of interest for information overload alleviation on the web, such as online retail platforms (\eg, Amazon, Alibaba)~\cite{2019online}, social networking applications (\eg, Facebook, Wechat) and online video sites (\eg, Youtube, Tiktok)~\cite{guo2020survey}. Among various techniques, Collaborative filtering (CF) has emerged as the foundation architecture for recommendation, such as conventional matrix factorization methods~\cite{mnih2008probabilistic,he2016fast}, and recent neural network-based models~\cite{he2017neural,wang2019unified,zou2020neural}. The key rationale behind the CF paradigm is to project users and items into low-dimensional latent representations, based on the observed user-item interactions~\cite{dong2017hybrid}.

With the advancement of graph neural networks (GNNs) in achieving unprecedented success for graph representations tasks, one recent promising research line leverages graph neural networks to exploit user-item interaction graph structure for collaborative filtering. Several recent efforts have explored deeply into GNNs to propagate embeddings with user-item interaction edges to capture CF signals among neighboring nodes in subgraphs, so as to further improve the user representation quality. Specifically, PinSage~\cite{ying2018graph} and NGCF~\cite{wang2019neural} employ GCN to propagate embeddings over the user-item interaction graph. Later on, LightGCN~\cite{he2020lightgcn} proposes to alleviate training difficulty by identifying the unnecessity of feature transformation in GNN-based recommender system. Additionally, inspired by the effectiveness of disentangled graph representation learning~\cite{ma2019disentangled}, DGCF~\cite{wang2020disentangled} disentangles latent user interests with intent-aware user-item graph structures for embedding users.

\begin{figure}
    \centering
    \includegraphics[width=\columnwidth]{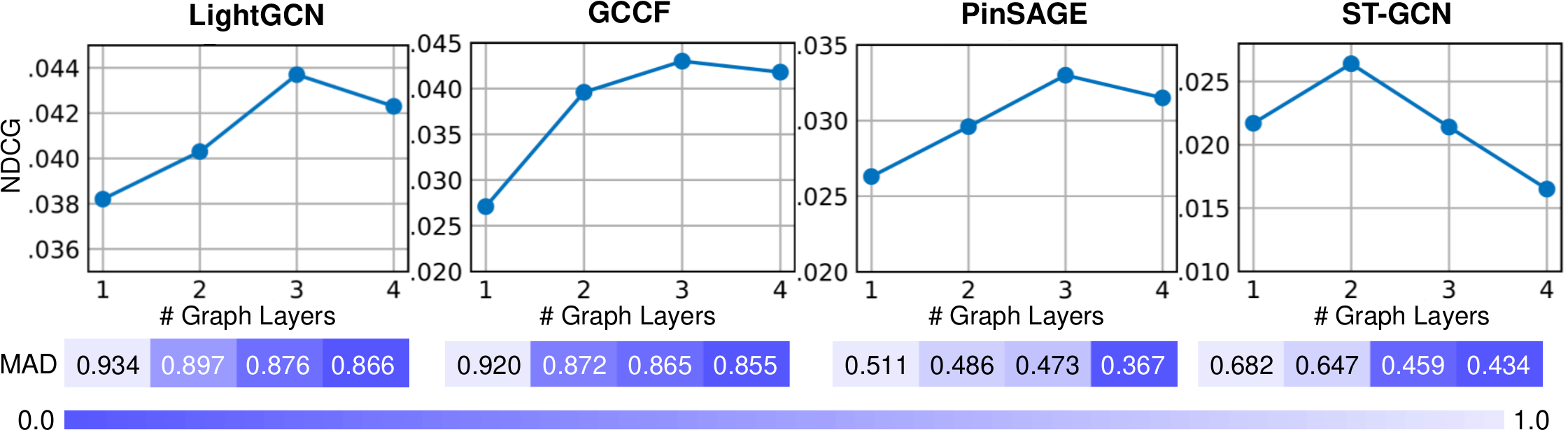}
    \vspace{-0.25in}
    \caption{i) The recommendation performance (measured by NDCG@20); and ii) The embedding smoothness (measured by MAD metric with heatmaps) of state-of-the-art graph-based collaborative filtering models (\ie, LightGCN~\cite{he2020lightgcn}, PinSage~\cite{ying2018graph}, GCCF~\cite{chen2020revisiting}, ST-GCN~\cite{zhang2019star}) on Yelp dataset.}
    \label{fig:over_smoonthing}
    \vspace{-0.25in}
\end{figure}

While the aforementioned graph-based Collaborative Filtering models have offered state-of-the-art performance for recommendation applications, two key issues remain less explored: 

i) \textbf{Over-Smoothing Collaborative Effects}. The graph neural CF architecture with deeper embedding propagation layers may result in indistinguishable user
vectors~\cite{zhou2020towards,min2020scattering}, which limits the representation quality of high-order collaborative relations. The over-mixing of information among connected users/items over their interaction subgraphs can involve harmful noise for representing unique user preference. The over-smoothing issues of graph-based CF paradigms (\eg, LightGCN~\cite{he2020lightgcn}, PinSage~\cite{ying2018graph}, ST-GCN~\cite{zhang2019star}, GCCF~\cite{chen2020revisiting}) are illustrated in Figure~\ref{fig:over_smoonthing}. In particular, we observe the larger smoothness degrees of the user representations encoded from different methods with the increase of graph propagation layers. Here, we adopt the quantitative metric Mean Average Distance (MAD)~\cite{chen2020measuring} to measure the graph smoothness by calculating the average distance among user embeddings learned by different methods. As shown in Figure~\ref{fig:over_smoonthing}, the over-smoothing issue limits the recommendation performance of graph neural CF frameworks due to the indistinguishable embeddings by stacking more graph aggregation layers. Hence, encoding distinguishable and informative user/item representations from the holistic interaction graph is of crucial importance for differentiating user preference.\\\vspace{-0.12in}

ii) \textbf{Supervision Interaction Data Scarcity and Noise}: Most graph-based collaborative filtering models belong to the supervised learning paradigm, in which the user representation process is supervised by user-item interaction signals. The effectiveness of such approaches largely relies on sufficient supervision signals (\ie, observed user-item interactions), and cannot learn quality user/item representations under the interaction label scarcity. However, the interaction data scarcity issue is ubiquitous in various practical recommendation scenarios~\cite{zheng2019deep}, since most users only sporadically interact with very limited number of items. The majority of collaborative filter-based recommender systems cannot perform well on long-tail items because of the data sparsity phenomenon and skewed interaction distribution. Additionally, most GNN-based recommendation methods design the message passing mechanism in which the embedding propagation is merely performed with the original graph neighbors. However, such explicit user-item relations (\eg, clicks or views) always bring noisy information (\eg, user misclick behaviors) to the constructed interaction graph~\cite{zhao2021heterogeneous}. For example, user's real interest is likely to be confounded after his/her page views on a lot of irrelevant items~\cite{wu2021self}.\\\vspace{-0.12in}

\noindent \textbf{Presented Work}. In view of the aforementioned limitations and challenges, we develop a hypergraph contrastive learning framework with self-augmentation for GNN-based collaborative relation modeling, by proposing a new recommendation model--\model. In particular, we leverage hypergraph learning architecture to supplement the encoding of graph-based CF paradigm with global collaborative effects based on the low-rank structure learning. While several recent studies use hypergraph to encode the high-order relations in recommendation, most of them rely on the original connections among users or items to generate their hypergraph connections. For example, DHCF~\cite{ji2020dual} constructs the hypergraph between users and items based on their observed interactions. MHCN~\cite{yu2021self} pre-defines the motif-based social connections for hypergraph generation. Nevertheless, such observed connections are inevitably noisy or incomplete in practical scenarios~\cite{zhao2021heterogeneous}. To mitigate this issue, in our \model, the parameterized hypergraph-guided user dependent structures and the original interaction graph encoder are jointly learned towards better user preference representation.

Furthermore, we design relational learning augmentation schema based on the dual-graph contrastive learning. Inspired by the strength of contrastive learning in visual and language data representation~\cite{chen2020simple,stojanovski2020contracat}, there exist some recent self-supervised collaborative filtering models (\eg, SGL~\cite{wu2021self}, SLRec~\cite{yao2021self}) which augment the original interaction data with randomly node and edge dropout, or mask operations. However, this might drop some essential information (\eg, interactions between users and their interested items), which hinders the encoding of user real interests. To fill this gap, we treat the explicit interaction graph and the learned implicit hypergraph structure as two contrastive views, without the random noise perturbation based on the pre-defined operators. We also provide theoretical justification behind our model design from the perspective of improving model optimization with better gradients. Empirical results show that \model\ consistently improves the performance over state-of-the-art recommendation models, \eg, average relative improvement of 24.7\% over LightGCN, 33.8\% over DHCF, 20.2\% over MHCN and 14.8\% over SGL in terms of NDCG@20.

In summary, this paper makes the following contributions:
\begin{itemize}[leftmargin=*]

\item We introduce a generic self-supervised recommendation framework for enhancing the robustness of graph collaborative filtering paradigm, through distilling self-augmented contrastive views between the local and global modeling of collaborative effects.

\item We present \model\ which innovatively integrates the global hypergraph structure learning with local collaborative relation encoder, to cooperatively supervise each other. The designed hypergraph contrastive learning schema empowers the graph neural CF paradigm, to capture the intrinsic and implicit dependencies among users and items with effective instance discrimination.

\item Systematic experimental studies are conducted to evaluate the performance of our \model\ model and 15 various baselines on several benchmark datasets. Further ablation study is provided to investigate the effect of key modules in \model. 


\end{itemize}

\section{Preliminaries and Related Work}
\label{sec:relate}


\subsection{Collaborative Filtering Learning Paradigm}
We let $\mathcal{U}$ = $\{u_1,...,u_i,...,u_I\}$ ($|\mathcal{U}| = I$) and $\mathcal{V}$ = $\{v_1,...,v_j,...,v_J\}$ ($|\mathcal{V}| = J$) represent the set of users and items, respectively. The interaction matrix $\mathcal{A} \in \mathbb{R}^{I\times J}$ indicates the implicit relationships between each user in $\mathcal{U}$ and his/her consumed items. Each entry $\mathcal{A}_{i,j}$ in $\mathcal{A}$ will be set as 1 if user $u_i$ has adopted item $v_j$ before and $\mathcal{A}_{i,j}=0$ otherwise. The objective of CF task is to forecast the unobserved user-item interactions with the encoded corresponding representations. The assumption of CF paradigm lies in that behaviorally similar users are more likely to share similar interests.



Many existing CF approaches are designed with the various embedding functions to generate vectorized representations of users and items. Then, the similarity matching function is introduced to estimate the relevance score between user $u_i$ and the candidate item $v_j$. Following this paradigm, NCF~\cite{he2017neural} and DMF~\cite{xue2017deep} replaces the inner-product with Multilayer Perceptron to reconstruct the ground truth user-item interactions. Furthermore, to transform users and items into the latent embeddings, Autoencoder has been utilized as the embedding function with the behavior reconstruction objective, such as AutoRec~\cite{sedhain2015autorec} and CDAE~\cite{wu2016collaborative}.


\vspace{-0.03in}
\subsection{Graph-based Recommender Systems.}
To capture high-order collaborative signals, one prominent direction explores user-item relations based on multi-hop interaction topological structures with graph oriented approaches~\cite{wang2019neural,huang2021graph,ying2018graph,xia2021graph}. For example, NGCF~\cite{wang2019neural} and PinSage~\cite{ying2018graph} have demonstrated the importance of high-order connectivity between users and items for collaborative filtering. To further improve the message passing process in graph convolutional framework, LightGCN~\cite{he2020lightgcn} proposes to omit the non-linear transformation during propagation, and uses the sum-based pooling operation for neighborhood aggregation.

Some recent studies also follow the graph-structured information propagation rule to refine user/item embeddings, with various neighborhood aggregation functions~\cite{huang2021recent}. For example, learning disentangled or behavior-aware user representations is proposed to improve CF paradigm, \eg, DGCF~\cite{wang2020disentangled}, MacridVAE~\cite{ma2019learning} and MBGMN~\cite{xia2021graph}. The hyperbolic embedding space is adopted to encode high-order information from neighboring users/items in~\cite{sun2021hgcf}.\vspace{-0.1in}


\vspace{-0.03in}
\subsection{Hypergraph Learning for Recommendation}
Inspired by the generalization ability of hypergraph in modeling complex high-order dependencies~\cite{feng2019hypergraph,gao2020hypergraph,he2021click}, some recently developed recommender systems are empowered to capture interaction patterns with the constructed hypergraph structures and uniform node-hyperedge connections, like HyRec~\cite{wang2020next}, DHCF~\cite{ji2020dual} and MHCN~\cite{yu2021self}. For instance, HyRec attempts to propagate information among multiple items by considering users as hyperedges. DHCF models the hybrid multi-order correlations between users and items based on the constructed hypergraph structures. Different from them, our \model\ framework designs a learnable hypergraph structure encoder, which not only improves the discrimination capability of CF representation paradigm, but also preserves the personalized global collaborative relationships across users and items in recommender systems. The presented hypergraph dependency structure learning method brings advantages in automatically distilling inter-dependency to enhance discrimination ability of user and item representations by addressing the graph over-smoothing issue.\vspace{-0.1in}




\vspace{-0.03in}
\subsection{Contrastive Representation Learning}
Contrastive learning has become an effective self-supervised framework, to capture the feature representation consistency under different views~\cite{oord2018representation,wei2022contrastive}. It has achieved promising performance in various domains, such as visual data representation~\cite{chen2020simple,peng2021self}, language data understanding~\cite{stojanovski2020contracat,cao2021grammatical}, graph representation learning~\cite{qiu2020gcc,zhu2021graph} and recommender systems~\cite{wu2021self,long2021social,yu2021self,kgcontrastive2022}. These contrastive learning approaches seek the exploration of data- or task-specific augmentations with auxiliary signals. To tackle the challenge of insufficient supervision labels in recommendation, we propose a new self-supervised recommendation framework to supplement the encoding of collaborative effects with explicitly local-global inter-dependency modeling, under a hypergraph learning schema.
\section{Methodology}
\label{sec:solution}

\begin{figure}
    \centering
    \includegraphics[width=0.96\columnwidth]{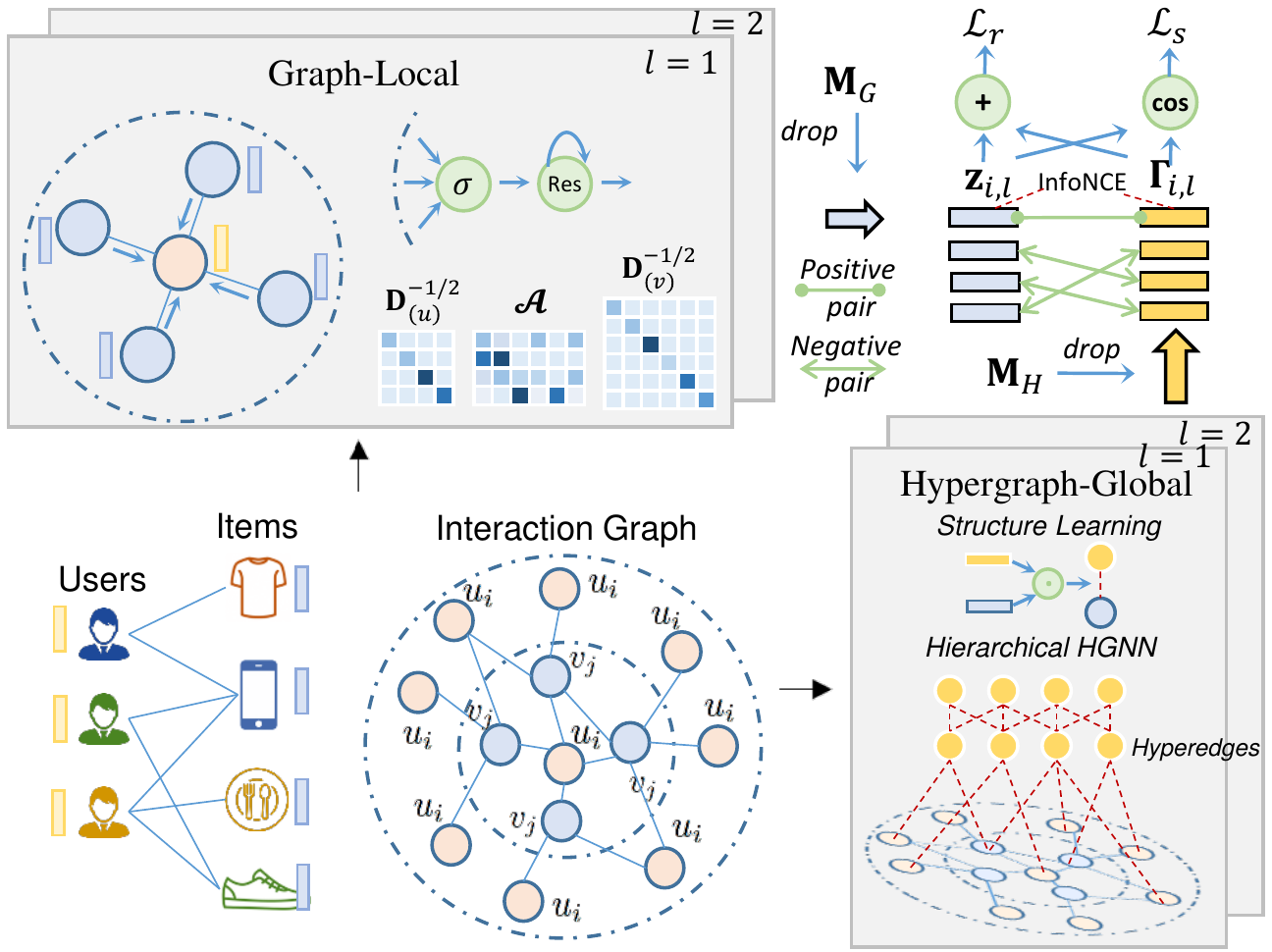}
    \vspace{-0.1in}
    \caption{The illustration of \model's overall architecture. i) In the Graph-Local component, the explicit local collaborative similarities are encoded with the embedding aggregation function: $\textbf{z}_i^{(u)} = \sigma(\bar{\mathcal{A}}_{i,*} \cdot \textbf{E}^{(v)})$ and $\textbf{z}_j^{(v)} = \sigma(\bar{\mathcal{A}}_{*,j} \cdot \textbf{E}^{(u)})$. ii) In the Hypergraph-Global component, the global cross-user and cross-item dependencies are captured through the hypergraph message passing function: $\sigma(\mathcal{H}^{(u)}\cdot\mathcal{H}^{(u)\top} \cdot \textbf{E}^{(u)}_{l-1})$. iii) In the cross-view contrastive component, the self-supervision signals are distilled from both local and global perspectives with the representations $\textbf{z}_{i,l}$ and global $\mathbf{\Gamma}_{i,l}$.}
    \vspace{-0.15in}
    \label{fig:framework}
\end{figure}

In this section, we present our \model\ framework whose overall architecture is shown in Figure~\ref{fig:framework}. First, we leverage the graph-based message passing module as the encoder to capture the local collaborative similarities among users and items. Second, we propose a new hypergraph neural network with global dependency structure learning to comprehensively capture global collaborative effects for graph neural CF paradigm. Finally, a new hypergraph contrastive learning architecture is introduced with complementary self-distilling views (local and global collaborative relations).

\subsection{Local Collaborative Relation Encoding}
Following the common collaborative filtering paradigm, we first represent each user $u_i$ and item $v_j$ with the embedding vectors $\textbf{e}_{i}^{(u)} \in\mathbb{R}^d$ and $\textbf{e}_{j}^{(v)} \in\mathbb{R}^d$, respectively ($d$ denotes the embedding dimensionality). We further define $\textbf{E}^{(u)}\in\mathbb{R}^{I\times d}$ and $\textbf{E}^{(v)} \in \mathbb{R}^{J\times d}$ to represent the embeddings corresponding to users and items. Inspired by the effectiveness of simplified graph convolutional network in LightGCN~\cite{he2020lightgcn}, we design our local graph embedding propagation layer with the following form:
\begin{align}
    \textbf{z}_i^{(u)} = \sigma(\bar{\mathcal{A}}_{i,*} \cdot \textbf{E}^{(v)}), ~~~
    \textbf{z}_j^{(v)} = \sigma(\bar{\mathcal{A}}_{*,j} \cdot \textbf{E}^{(u)})
\end{align}
\noindent where $\textbf{z}_i^{(u)}, \textbf{z}_j^{(v)} \in \mathbb{R}^d$ represent the aggregated information from neighboring items/users to the centric node $u_i$ and $v_j$. $\sigma(\cdot)$ denotes the LeakyReLU activation function with 0.5 slope for negative inputs, which not only brings benefits to the gradient back-propagation, but also injects the non-linearity into the transformation. Here, $\bar{\mathcal{A}}\in\mathbb{R}^{I\times J}$ denotes the normalized adjacent matrix derived from the user-item interaction matrix $\mathcal{A}$ calculated as:
\begin{align}
    \bar{\mathcal{A}} = \textbf{D}^{-1/2}_{(u)} \cdot \mathcal{A} \cdot \textbf{D}^{-1/2}_{(v)},~~~\bar{\mathcal{A}}_{i,j} = \frac{\mathcal{A}_{i,j}}{\sqrt{|\mathcal{N}_i|\cdot|\mathcal{N}_j|}}
\end{align}
where $\textbf{D}_{(u)}\in\mathbb{R}^{I\times I}, \textbf{D}_{(v)}\in\mathbb{R}^{J\times J}$ are diagonal degree matrices. The neighboring items/users of user ($u_i$)/item ($v_j$) are denoted by $\mathcal{N}_i$ and $\mathcal{N}_j$, respectively.

By integrating multiple embedding propagation layers, we refine the user/item representations to aggregate local neighborhood information for contextual embedding generation. Suppose $\textbf{e}_{i,l}^{(u)}$ and $\textbf{e}_{j,l}^{(v)}$ represents the embedding of user $u_i$ and item $v_j$ at the $(l)$-th GNN layer. The message passing process from $(l-1)$-th layer to the $(l)$-th layer is formally defined as below:
\begin{align}
    \textbf{e}_{i,l}^{(u)} = \textbf{z}_{i,l}^{(u)} + \textbf{e}_{i, l-1}^{(u)}, ~~~ \textbf{e}_{j,l}^{(v)} = \textbf{z}_{j,l}^{(v)} + \textbf{e}_{j,l-1}^{(v)}
\end{align}
\noindent We apply the residual connections for self-information incorporation between the source and target node during the message aggregation across graph layers. This emphasizes the semantics of the centric node and alleviates the over-smoothing issue of GNN.

\begin{figure*}[t]
    \centering
    \includegraphics[width=2.0\columnwidth]{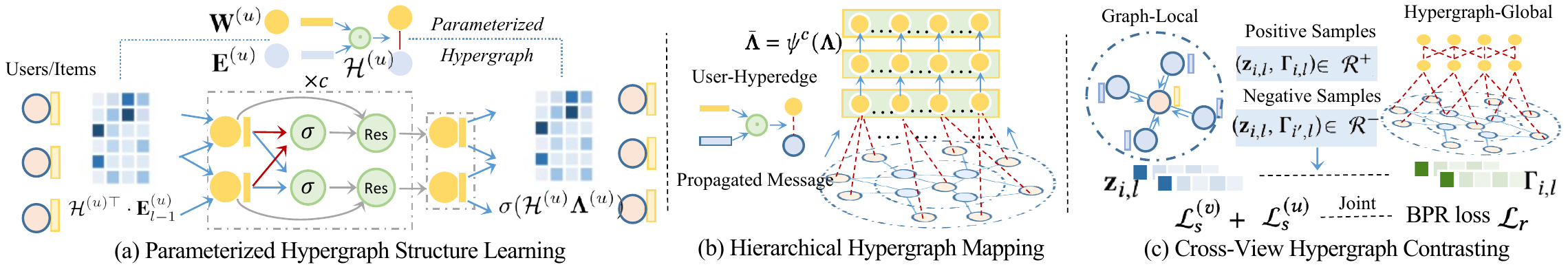}
    \vspace{-0.1in}
    \caption{The illustration of hypergraph-enhanced contrasting architecture built over the hypergraph structure learning. (a) The hypergraph structure is parameterized with $\mathcal{H}^{(u)}$ and $\mathcal{H}^{(v)}$. (b) $\psi^c(\mathbf{\Lambda})$ represents the hierarchically structured multiple hypergraph neural layers. (c) Contrastive objectives $\mathcal{L}_s^{(u)}$ and $\mathcal{L}_s^{(v)}$ are generated via local-global cross-view supervision. }
    \vspace{-0.1in}
    \label{fig:hypergraph_framework}
\end{figure*}

\subsection{Hypergraph Global Dependency Learning}
To inject global collaborative relations among users/items into the CF representation paradigm, we design a global dependency encoder via hypergraph structure learning.

\subsubsection{\bf Hypergraph Message Passing Paradigm.} Motivated by the strength of hypergraph for unifying nodes beyond pairwise relations~\cite{jiang2019dynamic}, we endow our \model\ to capture complex high-order relations under a deep hypergraph message passing architecture. Hypergraph consists of a set of vertices and hyperedges, in which each hyperedge can connect any number of vertices~\cite{chen2020neural}. In our hypergraph collaborative filtering scenario, we utilize hyperedges as intermediate hubs for global-aware information passing across users and items without the hop distance limit. The model structure of hypergraph message passing paradigm is illustrated in Figure~\ref{fig:hypergraph_framework}.

To be specific, we define the hypergraph dependency matrix for users and items as $\mathcal{H}^{(u)}\in\mathbb{R}^{I\times H}$ and $\mathcal{H}^{(v)}\in\mathbb{R}^{J\times H}$. Here, $H$ represent the number of hyperedges. We give a formal presentation of our hypergraph message passing as:
\begin{align}
    \mathbf{\Gamma}^{(u)}_l = \sigma(\mathcal{H}^{(u)}\mathbf{\Lambda}^{(u)})=\sigma(\mathcal{H}^{(u)}\cdot\mathcal{H}^{(u)\top} \cdot \textbf{E}^{(u)}_{l-1})
\end{align}
where $\mathbf{\Lambda}^{(u)}\in\mathbf{R}^{H\times d}$ denotes the hyperedge-specific embeddings for users, and $\sigma$ denotes the LeakyReLU mapping. $\mathbf{\Gamma}^{(u)}_l \in\mathbb{R}^{I\times d} $ represents the hyper embeddings of users ($u_i \in \mathcal{U}$) in hypergraph representation space under the $l$-th propagation layer. The hyper embeddings $\mathbf{\Gamma}^{(v)}_l$ of items ($v_j \in \mathcal{V}$) can be generated in an analogous way. Our hypergraph message extraction phase takes the user/item embeddings $\textbf{E}^{(u)}_{l-1}$, $\textbf{E}^{(v)}_{l-1}$ (refined from the graph local embedding propagation), and learnable hypergraph dependeny structural matrices $\mathcal{H}^{(u)}$, $\mathcal{H}^{(v)}$ (will be elaborated in the following subsection) as the computation input knowledge, to jointly preserve the local and global collaborative effects.\\\vspace{-0.12in}

\subsubsection{\bf Parameterized Hypergraph Structure Learning.}
To adaptively learn hypergraph-based dependent structures across users and items, we propose to paramerterize hypergraph dependency matrices $\mathcal{H}^{(u)}$, $\mathcal{H}^{(v)}$ which are jointly optimized along with the GNN-based CF architecture. Without loss of generality, we use $\mathcal{H}$ to denote user-side dependency matrix $\mathcal{H}^{(u)}$ or item-side dependency matrix $\mathcal{H}^{(v)}$ in the following sections for simplifying notations. Note that by obtaining a trainable hypergraph structural matrix $\mathcal{H}$, the node-wise global dependency can be derived with $\mathcal{H}\mathcal{H}^\top$.

However, learning dense hypergraph adjacent matrices $\mathcal{H}^{(u)}\in\mathbb{R}^{I\times H},\mathcal{H}^{(v)}\in\mathbb{R}^{J\times H}$ will enormously increase the size of model parameters with high computational cost (especially for the large number of hyperedges $H$). To tackle this challenge, we propose to parameterize the hypergraph structure matrices $\mathcal{H}^{(u)},\mathcal{H}^{(v)}$ into latent space in a low-rank manner to achieve model efficiency:
\begin{align}
    \mathcal{H}^{(u)} = \textbf{E}^{(u)}\cdot \textbf{W}^{(u)},~~~~
    \mathcal{H}^{(v)} = \textbf{E}^{(v)}\cdot \textbf{W}^{(v)}
\end{align}
\noindent where $\textbf{W}^{(u)},\textbf{W}^{(v)}\in\mathbb{R}^{d\times H}$ represents the learnable embedding matrices for user- and item-side hyperedges. $\textbf{E}^{(u)},\textbf{E}^{(v)}$ represents the embeddings of users and items. By doing so, the hypergraph structure learning component only takes extra $O(H\times d)$ time complexity, which is quite small as compared to the computational cost $O((I+J)\times d)$ for the main embedding space. In addition, we design the parameterized dependency structure encoder with the matrix decomposition on $\mathcal{H}$ in a low-rank manner, which further regularizes the latent structure learning for overfitting alleviation.

\subsubsection{\bf Hierarchical Hypergraph Mapping.}
Although the aforementioned hypergraph structure learning framework endows our \model\ with the capability of capturing global user- and item-wise collaborative relationships, we further supercharge our hypergraph neural architecture with a high-level of hyperedge-wise feature interaction. Towards this end, we augment our \model\ with multiple hypergraph neural layers by stacking different layers of hyperedges with size $H$. In particular, the set of foregoing hyperedges will serve as the first hypergraph layer, which will interact with deep hypergraph layers non-linearly. In form, the hyperedge embeddings of deep layers are derived with the encoding function $\psi(\cdot)$ as follows:
\begin{align}
    \bar{\mathbf{\Lambda}} = \psi^c(\mathbf{\Lambda}),~
    \psi(\textbf{X}) = \sigma(\textbf{V} \textbf{X}) + \textbf{X},~\mathbf{\Lambda}=\mathcal{H}^\top\textbf{E}
\end{align}
\noindent where $\mathbf{\Lambda}\in\mathbb{R}^{H\times d}$ denotes the hyperedge embeddings aggregated from the user/item representations $\textbf{E}\in\mathbb{R}^{K\times d}$ as well as the learned node-hyperedge structures $\mathcal{H}\in\mathbb{R}^{K\times H}$ ($K=I$ or $J$). Here, $c$ denotes the number of hypergraph embedding layers. $\textbf{V}\in\mathbb{R}^{H\times H}$ is a trainable parametric matrix for embedding projection. We adopt LeakyReLU as the activation function $\sigma(\cdot)$ to handle non-linearities. The residual connection operation is applied to the embedding generation across hypergraph layers. After the hierarchical hypergraph mapping, we refine the user/item representations:
\begin{align}
    \mathbf{\Gamma}_l=\sigma(\mathcal{H} \cdot \bar{\mathbf{\Lambda}}) = \sigma(\mathcal{H}\cdot \psi^c(\mathcal{H}^\top\cdot\textbf{E}))
\end{align}

\subsection{Multi-Order Aggregation and Prediction}
To integrate the local dependency encoding with the global collaborative relation modeling, we iteratively perform the graph local embedding propagation and hypergraph information aggregation:
\begin{align}
    \nonumber
    \textbf{e}_{i,l}^{(u)}&=\textbf{z}_{i,l}^{(u)}+ \mathbf{\Gamma}_{i,l}^{(u)} + \textbf{e}_{i,l-1}^{(u)},~\textbf{e}_{j,l}^{(v)}=\textbf{z}_{j,l}^{(v)}+ \mathbf{\Gamma}_{j,l}^{(v)} + \textbf{e}_{j,l-1}^{(v)}
\end{align}
\noindent where $\textbf{z}_{i,l}^{(u)}$ represents the $l$-th order embedding of user $u_i$ aggregated from his/her locally connected neighbors. $\mathbf{\Gamma}_{i,l}^{(u)}$ denotes the fused $l$-th order user representation through hyperedges. We further apply the residual operations for the embedding aggregation. The multi-order user/item embeddings are generated with the element-wise embedding summation, and the inner product of them is utilized to estimate the interaction preference score between user $u_i$ and item $v_j$, formally is shown as follows:
\begin{align}
    \mathbf{\Psi}_i^{(u)}=\sum_{l=0}^L \textbf{E}_{i,l}^{(u)},~ \mathbf{\Psi}_j^{(v)}=\sum_{l=0}^L \textbf{E}_{j,l}^{(v)},~ 
    \text{Pr}_{i,j}=\mathbf{\Psi}_i^{(u)\top}\mathbf{\Psi}_j^{(v)}
\end{align}
Given these notations, we define our pair-wise marginal loss as:
\begin{align}
    \mathcal{L}_r=\sum_{i=0}^I\sum_{s=1}^S \max(0, 1-\text{Pr}_{i,p_s}+\text{Pr}_{i,n_s})
\end{align}
\noindent For each user $u_i$, we respectively sample $S$ positive (indexed by $p_s$) and negative (indexed by $n_s$) instances from their observed and non-interacted items.

\subsection{Hypergraph-enhanced Contrastive Learning}
This section describes how we enable our \model\ with the cross-view collaborative supervision under a hypergraph neural architecture, to augment the user representation with sparse supervision signals. 

\subsubsection{\bf Hypergraph-guided Contrasting.}
We design our contrastive learning component by maximizing the agreement between the explicit user-item interactive relationships and the implicit hypergraph-based dependency. In particular, we generate two representation views as i) local collaborative relation encoding over the user-item interaction graph, and ii) global hypergraph structure learning among users/items. Such contrastive learning leverages the user and item self-discrimination, to offer auxiliary supervision signals from the local and global representation space.

\subsubsection{\bf Cross-View Collaborative Supervision.}
We take different views of the same user/item as the positive pairs ($\textbf{z}_{i,l}$, $\mathbf{\Gamma}_{i,l}$), and treat views of different users/items as negative pairs. By doing so, our model learns discriminative representations by contrasting the generated positive and negative instances. We formally define our contrastive loss for user representations with the InfoNCE~\cite{oord2018representation} as:
\begin{align}
    \mathcal{L}_s^{(u)} = \sum_{i=0}^I\sum_{l=0}^L -\log \frac{\exp(s(\textbf{z}_{i,l}^{(u)}, \mathbf{\Gamma}_{i,l}^{(u)})/\tau)}{\sum_{i'=0}^I \exp(s({\textbf{z}_{i,l}^{(u)}, \mathbf{\Gamma}_{i',l}^{(u)})/\tau })}
\end{align}
\noindent where $s(\cdot)$ denotes the cosine similarity function and $\tau$ denotes the tunable temperature hyperparameter to adjust the scale for softmax. We perform the contrastive learning between the local user embedding ($\textbf{z}_{i,l}^{(u)}$) and global hypergraph-guided representation $\mathbf{\Gamma}_{i,l}^{(u)}$. This allows the local and global dependency views to collaboratively supervise each other, which enhances the user representation.

\subsubsection{\bf Data Augmentation on Graph Structure.}
To further alleviate the overfitting issue during the cross-view contrastive learning process, we design edge dropout operator over both the user-item interaction graph and the learned hypergraph structure as:
\begin{align}
    \bar{\mathcal{A}} := \textbf{M}_G \circ \bar{\mathcal{A}};~~~~~\mathcal{H} := \textbf{M}_H \circ \mathcal{H}
\end{align}
\noindent where $:=$ denotes the assignment operator and $\circ$ denotes the element-wise multiplication. $\textbf{M}_G\in\mathbb{R}^{I\times J}$ and $\textbf{M}_H\in\mathbb{R}^{J\times H}$ are binary mask matrices with dropout probability $\mu$. We integrate our hypergraph contrastive loss with our CF loss into a unified objective as:
\begin{align}
    \mathcal{L}=\mathcal{L}_r+\lambda_1\cdot(\mathcal{L}_s^{(u)}+\mathcal{L}_s^{(v)})+\lambda_2\cdot\|\mathbf{\Theta}\|_{\text{F}}^2
\end{align}
\noindent We minimize $\mathcal{L}$ using Adam optimizer. The weight-decay regularization term is applied over parameters $\mathbf{\Theta}$.

\subsection{In-Depth Analysis of \model}
This section provides further analysis of our \model\ model with the theoretical discussion and time complexity analysis.

\subsubsection{\bf Theoretical Analysis of \model} Our hypergraph contrastive learning enhances the discrimination ability of graph-based CF representation paradigm by generating adaptive gradients from hard negative samples. Specifically, the influences of negative samples over the learned embeddings under the self-supervised learning architecture can be quantified as:
\begin{align}
    c(i')=\left(\bar{\mathbf{\Gamma}}_{i',l}^{(u)}-(\bar{\textbf{z}}_{i,l}^{(u)\top} \bar{\mathbf{\Gamma}}_{i',l}^{(u)})\bar{\textbf{z}}_{i,l}^{(u)}\right) \cdot \frac{\exp(\bar{\textbf{z}}_{i,l}^{(u)\top} \bar{\mathbf{\Gamma}}_{i,l}^{(u)}/\tau)}{\sum_{i'} \exp(\bar{\textbf{z}}_{i,l}^{(u)\top} \bar{\mathbf{\Gamma}}_{i',l}^{(u)}/\tau)} 
\end{align}
\noindent where $c(i')$ denotes the gradients related to negative sample $i'$, $\bar{\mathbf{\Gamma}}_{i',l}^{(u)}=\mathbf{\Gamma}_{i',l}^{(u)} / \|\mathbf{\Gamma}_{i',l}^{(u)}\|_2$ and $\bar{\textbf{z}}_{i,l}^{(u)}={\textbf{z}}_{i,l}^{(u)} / \|{\textbf{z}}_{i,l}^{(u)}\|_2$ are the normalized embeddings. By further analyzing the norm of $c(i')$, we can find it proportional to a unique function as:
\begin{align}
    \|c(i')\|_2 \propto \sqrt{1-x^2}\cdot\exp(\frac{x}{\tau})
\end{align}
\noindent where $x=\bar{\textbf{z}}_{i,l}^{(u)\top} \bar{\mathbf{\Gamma}}_{i',l}^{(u)}$ is the cosine similarity between the embeddings from two contrastive views. As the similarity score increases, $\|c(i')\|_2$ increases dramatically, under proper $\tau$ settings. Inspired by the recent studies in~\cite{khosla2020supervised,wu2021self}, our contrastive learning emphasizes the gradients from hard negative samples that are more similar to the positive ones, which improves the model training. In our hypergraph structure learning, we endow the CF paradigm with the power to capture the global user dependency without the limitation of observed user-item connections. This also allows the robust gradient updates of different user representations to influence with each other based on their latent global relatedness.

\subsubsection{\bf Model Complexity Analyses.}
The graph local collaborative relation encoding takes $O(L\times |\mathcal{A}|\times d)$ complexity, where $L$ denotes the number of graph neural layers. $|\mathcal{A}|$ is the number of edges in user-item interaction graph. Additionally, the hypergraph message passing schema takes $O(L\times(I+J)\times H\times d)$ complexity with the node-hyperedge information propagation. In our contrastive learning paradigm, the cost is $O(L\times B\times (I+J)\times d)$, owing to our designed low-rank-based hypergraph structure learning. Here, $B$ denotes the number of users/items included in a single batch. For the memory cost, \model\ only involves $O(H\times d)$ extra parameters in the hypergraph structure learning, as compared to existing GNN-based collaborative filtering models.
\section{Evaluation}
\label{sec:eval}


Our experiments aim to answer the research questions as follows:

\begin{itemize}[leftmargin=*]

\item \textbf{RQ1}: What is the performance of our \model\ as compared to various state-of-the-art recommender systems?

\item \textbf{RQ2}: How does the hypergraph structure learning and cross-view contrastive learning contribute to the performance?

\item \textbf{RQ3}: How does \model\ perform in alleviating data sparsity?

\item \textbf{RQ4}: How do the key hyperparameters influence the performance of the proposed \model\ framework?

\item \textbf{RQ5}: How does the hypergraph-enhanced global dependency modeling benefit the model interpretation?

\end{itemize}

\subsection{Experimental Settings}

\begin{table}[]
    \centering
    \small
    \caption{Statistics of the experimental datasets.}
    \vspace{-0.15in}
    \begin{tabular}{ccccc}
        \toprule
        Dataset & User \# & Item \# & Interaction \# & Density\\
        \midrule
        Yelp & 29601 & 24734 & 1517326 & $2.1e^{-3}$\\
        Movielens & 69878 & 10196 & 9988816 & $1.4e^{-2}$\\
        Amazon-book & 78578 & 77801 & 3190224 & $5.2e^{-4}$\\
        \bottomrule
    \end{tabular}
    \vspace{-0.15in}
    \label{tab:data}
\end{table}

\begin{table*}[t]
\caption{Performance comparison on Yelp, MovieLens, Amazon datasets in terms of \textit{Recall} and \textit{NDCG}.}
\vspace{-0.15in}
\centering
\footnotesize
\setlength{\tabcolsep}{1mm}
\begin{tabular}{|c|c|c|c|c|c|c|c|c|c|c|c|c|c|c|c|c|c|c|}
\hline
Data & Metric & BiasMF & NCF & AutoR & GCMC & PinSage & NGCF & STGCN & LightGCN & GCCF & DGCF & HyRec & DHCF & MHCN & SLRec & SGL & \emph{\model} & p-val.\\
\hline
\multirow{4}{*}{Yelp}
&Recall@20 & 0.0190 & 0.0252 & 0.0259 &  0.0266 & 0.0345 & 0.0294 & 0.0309 & 0.0482 & 0.0462 & 0.0466 & 0.0472 & 0.0449 & 0.0503 & 0.0476 & 0.0526 & \textbf{0.0607} & $2\times 10^{-6}$\\
&NDCG@20 & 0.0161 & 0.0202 & 0.0210 & 0.0251 & 0.0288 & 0.0243 & 0.0262 & 0.0409 & 0.0398 & 0.0395 & 0.0395 & 0.0381 & 0.0424 & 0.0398 & 0.0444 & \textbf{0.0510} & $5\times 10^{-8}$ \\
\cline{2-19}
&Recall@40 & 0.0371 & 0.0487 & 0.0504 & 0.0585 & 0.0599 & 0.0522 & 0.0504 & 0.0803 & 0.0760 & 0.0774 & 0.0791 & 0.0751 & 0.0826 & 0.0821 & 0.0869 & \textbf{0.1007} & $3\times 10^{-6}$\\
&NDCG@40 & 0.0227 & 0.0289 & 0.0301 & 0.0373 & 0.0385 & 0.0330 & 0.0332 & 0.0527 & 0.0508 & 0.0511 & 0.0522 & 0.0493 & 0.0544 & 0.0541 & 0.0571 & \textbf{0.0658} & $4\times 10^{-7}$\\
\hline
\multirow{4}{*}{MLens}
&Recall@20 & 0.0878 & 0.1097 & 0.1230 & 0.1411 & 0.1706 & 0.1611 & 0.1298 & 0.1789 & 0.1742 & 0.1763 & 0.1801 & 0.1363 & 0.1497 & 0.1758 & 0.1833 & \textbf{0.2048} & $3\times 10^{-3}$\\
&NDCG@20 & 0.1197 & 0.1297 & 0.1667 & 0.1731 & 0.2108 & 0.1961 & 0.1639 & 0.2128 & 0.2109 & 0.2101 & 0.2178 & 0.1726 & 0.1814 & 0.2003 & 0.2205 & \textbf{0.2467} & $4\times 10^{-7}$\\
\cline{2-19}
&Recall@40 & 0.1753 & 0.1634 & 0.1908 & 0.2106 & 0.2724 & 0.2594 & 0.2006 & 0.2650 & 0.2606 & 0.2681 & 0.2685 & 0.2171 & 0.2250 & 0.2633 & 0.2768 & \textbf{0.3081} & $8\times 10^{-6}$\\
&NDCG@40 & 0.1381 & 0.1427 & 0.1785 & 0.1865 & 0.2362 & 0.2225 & 0.1782 & 0.2322 & 0.2331 & 0.2319 & 0.2340 & 0.1901 & 0.1962 & 0.2360 & 0.2426 & \textbf{0.2717} & $7\times 10^{-7}$\\
\hline
\multirow{4}{*}{Amazon}
&Recall@20 & 0.0093 & 0.0142 & 0.0131 & 0.0103 & 0.0213 & 0.0222 & 0.0192 & 0.0319 & 0.0317 & 0.0211 & 0.0302 & 0.0280 & 0.0296 & 0.0285 & 0.0327 & \textbf{0.0344} & $2\times 10^{-4}$\\
&NDCG@20 & 0.0049 & 0.0085 & 0.0099 & 0.0074 & 0.0158 & 0.0160 & 0.0144 & 0.0236 & 0.0243 & 0.0154 & 0.0225 & 0.0202 & 0.0219 & 0.0238 & 0.0249 & \textbf{0.0258} & $5\times 10^{-3}$\\
\cline{2-19}
&Recall@40 & 0.0196 & 0.0223 & 0.0202 & 0.0217 & 0.0366 & 0.0376 & 0.0312 & 0.0499 & 0.0483 & 0.0351 & 0.0432 & 0.0471 & 0.0489 & 0.0463 & 0.0531 & \textbf{0.0561} & $2\times 10^{-4}$\\
&NDCG@40 & 0.0103 & 0.0133 & 0.0123 & 0.0124 & 0.0210 & 0.0213 & 0.0184 & 0.0290 & 0.0285 & 0.0201 & 0.0246 & 0.0272 & 0.0284 & 0.0314 & 0.0312 & \textbf{0.0330} & $2\times 10^{-3}$\\
\hline
\end{tabular}
\vspace{-0.05in}
\label{tab:overall_performance}
\end{table*}

\subsubsection{\bf Evaluation Datasets} We conduct experiments on three public datasets. \textbf{Yelp}: this dataset has been widely adopted for evaluating recommender systems with the task of business venue recommendation. \textbf{MovieLens}: it is a movie recommendation dataset. We follow the same processing rubric in~\cite{he2017neural} using the 10-core setting by keeping users and items with at least 10 interactions. \textbf{Amazon-book}: this dataset records user ratings on products with book category on Amazon with the 20-core setting. Table~\ref{tab:data} presents the statistics of all datasets. Experimented datasets and codes have been released (refer to the link provided in the abstract). \\\vspace{-0.15in}


\subsubsection{\bf Evaluation Protocols and Metrics}
Following the same data partition rubrics in most recent graph CF models~\cite{yao2021self,wang2019neural}, we generate our training, validation and test set with the ratio of 7:1:2. To mitigate the sampling bias, we evaluate the prediction accuracy using the all-ranking protocol~\cite{he2020lightgcn} over all items. We adopt \textit{Recall@N} and \textit{Normalized Discounted Cumulative Gain (NDCG)@N} as the metrics, which are widely used in recommendation~\cite{wang2019neural,xia2021knowledge}.

\subsubsection{\bf Baseline Methods.} We compare \model\ with 15 baselines covering various recommendation paradigms.

\noindent \textbf{Conventional Matrix Factorization Approach.}\vspace{-0.05in}
\begin{itemize}[leftmargin=*]
\item \textbf{BiasMF}~\cite{koren2009matrix}. It is a widely adopted baseline which is developed over the matrix factorization with user and item bias.
\end{itemize}

\noindent \textbf{MLP-enhanced Collaborative Filtering.}\vspace{-0.05in}
\begin{itemize}[leftmargin=*]
\item \textbf{NCF}~\cite{he2017neural}. It is neural CF model which encodes the non-linear feature interactions with multiple hidden layers. In particular, we use two hidden layers with the same embedding dimensionality.
\end{itemize}

\noindent \textbf{Autoencoder-based Collaborative Filtering Model.}\vspace{-0.05in}
\begin{itemize}[leftmargin=*]
\item \textbf{AutoR}~\cite{sedhain2015autorec}. It utilizes Autoencoder as the embedding projection function with the reconstruction objective to generate latent representations for observed user-item interactions.
\end{itemize}

\noindent \textbf{GNN-based Collaborative Filtering Frameworks.}\vspace{-0.05in}
\begin{itemize}[leftmargin=*]

\item \textbf{GC-MC}~\cite{berg2017graph}: The graph convolutional operations are applied to capture user-item dependency based on neighbor relationships.

\item \textbf{PinSage}~\cite{ying2018graph}: In the message passing of PinSage, the neighbor relations are generated based on the random walk sampling strategy. It generates negative instances using the PageRank score.

\item \textbf{NGCF}~\cite{wang2019neural}: this approach designs graph embedding propagation layers to generate user/item representations, by aggregating feature embeddings with high-order connection information.

\item \textbf{ST-GCN}~\cite{zhang2019star}: it supplements the GCN-based with the reconstruction task of masked node embeddings to address the limitation of label leakage issue. ST-GCN integrates graph convolutional encoder-decoders with intermediate supervision.

\item \textbf{LightGCN}~\cite{he2020lightgcn}: it proposes to simply the burdensome NGCF framework by removing the non-linear projection and embedding transformation during the message passing. The sum-based pooling is used for user representation generation.

\item \textbf{GCCF}~\cite{chen2020revisiting}: it enhances the graph neural network-based CF from two perspectives: i) omitting the non-linear transformation; ii) incorporating the residual network structure.

\end{itemize}


\noindent \textbf{Disentangled Graph Learning for Recommendation.}\vspace{-0.05in}
\begin{itemize}[leftmargin=*]
\item \textbf{DGCF}~\cite{wang2020disentangled}. it investigates the fine-grained user intention to enhance the representation ability of CF framework by disentangling multiple latent factors for user representation.
\end{itemize}

\noindent \textbf{Recommendation with Hypergraph Neural Networks.}\vspace{-0.05in}
\begin{itemize}[leftmargin=*]
\item \textbf{HyRec}~\cite{wang2020next}. It leverages the hypergraph structure to model relationships between user and his/her interacted items, by considering multi-order information in a dynamic environment.
\item \textbf{DHCF}~\cite{ji2020dual}. The new developed jump hypergraph convolution is introduced into the dual-channel CF to perform message passing with prior information. The divide-and-conquer scheme is used for dual-channel learning.
\end{itemize}

\noindent \textbf{Self-Supervised Learning for Recommendation.}\vspace{-0.05in}
\begin{itemize}[leftmargin=*]
\item \textbf{MHCN}~\cite{yu2021self}. It enhances the graph neural network-based recommendation with the self-supervision signals generated from the graph informax network, by maximizing the mutual information between node embedding and graph-level representation.

\item \textbf{SGL}~\cite{wu2021self}. This state-of-the-art self-supervised graph learning method directly changes the structures of user-item interaction graphs for data augmentation with i) probability-based node and edge dropout operations; ii) random walk-based sampling.

\item \textbf{SLRec}~\cite{yao2021self}. In this method, the feature correlations are considered as the additional regularization for improving recommender systems with a multi-task self-supervised learning approach.
\end{itemize}

\vspace{-0.1in}
\subsubsection{\bf Hyperparameter Settings}
We use Adam optimizer with the learning rate of $1e^{-3}$ and 0.96 decay ratio for model inference. The hidden state dimensionality is configured as 32. We stack two propagation layers for graph local embedding propagation. In our hypergraph learning, the number of hyperedges and hierarchical hypergraph mapping layers are set as 128 and 3, respectively. The batch size and dropout ratio are selected from $\{64, 128, 256, 512\}$ and $\{0.25, 0.5, 0.75\}$, respectively. The regularization weight $\lambda_1$ and $\lambda_2$ are tuned from the ragne $\{1e^{-5},1e^{-4}, 1e^{-3}, 1e^{-2}\}$ for loss balance. The temperature parameter $\tau$ is searched from the range \{0.1, 0.3, 1, 3, 10\} to control the strength of gradients in our contrastive learning. The effect of $\tau$ is further studied in the following subsection.


\vspace{-0.05in}
\subsection{Performance Validation (RQ1)}

In this section, we analyze the evaluation results and summarize the following observations and conclusions.

\begin{itemize}[leftmargin=*]

\item \textbf{Overall Performance Validation}. As the experimental result shown in Table~\ref{tab:overall_performance}, our \model\ consistently outperforms all baselines across different datasets in terms of all evaluation metrics. This observation validates the superiority of our \model\ method, which can be attributed to: i) By uncovering latent global collaborative effects, \model\ can not only model the holistic dependencies among users, but also preserves individual user interaction patterns with better discrimination ability. ii) Benefiting from our hypergraph contrastive learning schema, \model\ fulfills the effective self-data augmentation for sparse interactions, with cross-view (from locally to globally) supervision signals.\\\vspace{-0.12in}

\item \textbf{Superiority of Hypergraph Structure Learning}. Additionally, hypergraph recommendation methods (\ie, DHCF and HyRec) outperform most of GNN-based CF models (\eg, PinSage, NGCF, ST-GCN), suggesting the effectiveness of modeling high-order collaborative effects under hypergraph architecture. The significant improvement of our \model\ over competitive hypergraph recommender systems, further indicates that our designed learnable hypergraph neural structure paradigm is good at i) adaptively fulfilling the modeling of global collaborative dependencies among users and items; ii) alleviating the over-smoothing effect among neighboring node embeddings in GNN-based CF models. Specifically, representations on user preference are refined by both local connected users/items and global dependent users. \\\vspace{-0.12in}

\item \textbf{Effectiveness of Hypergraph Contrastive Learning}. With the analysis of \model\ across different datasets, we notice that the performance gap with baselines on Amazon data is larger than other datasets. This observation validates the effectiveness of \model\ in handing the sparse user-item interactions. Furthermore, compared with state-of-the-art self-supervised learning recommendation models (\ie, MHCN, SGL, SLRec), \model\ consistently achieves better performance. Specifically, SGL uses the probability-based randomly masking operations to generate contrastive views, which may dropout important supervision interaction labels. Furthermore, following the generative self-supervised frameworks, MHCN and SLRec attempt to integrate additional learning tasks with the main recommendation objective. However, the incorporated self-supervised learning objective may not be able to adapt well on the recommendation loss optimization, which significantly hurts the representation ability of recommender systems. This justifies the superiority of our hypergraph contrastive learning paradigm, via effectively integrating hypergraph strucutre learning with the cross-view contrastive self-supervised signals for collaborative filtering.


\end{itemize}

\vspace{-0.1in}
\subsection{Ablation Study of \model\ (RQ2)}
We explore the component effects of \model\ from: i) global hypergraph structure learning; ii) cross-view contrastive self-augmentation.

\begin{itemize}[leftmargin=*]

\item \textbf{Effect of Hypergraph Structure Learning}.
We investigate the importance of hypergraph learning for contributing the performance improvement, by generating two variants by: (1) removing the hierarchical hypergraph mapping for hyperedge-wise feature interaction, termed as \emph{-HHM}; (2) disabling the low-rank hypergraph dependency encoding, termed as \emph{-LowR}.\\\vspace{-0.12in}

\textbf{Results}. We report the evaluation results in Table~\ref{tab:module_ablation}. Without the exploration of hierarchically structured hypergraph mapping, \emph{-HHM} downgrades the recommendation accuracy. This observation justifies the rationality of enabling the hierarchical non-linear feature interaction through deep hypergraph neural layers. Additionally, while \emph{-LowR} preserves the multi-layer hypergraph structures, it directly learns a $\mathbb{R}^{I\times H}$ transformation matrix with larger parameter size, and thus may lead to the overfitting. With the parameterized hypergraph structure learning in a low-rank manner, we not only simplify the model size but also alleviate the overfitting via effective global message passing. \\\vspace{-0.12in}

\item \textbf{Effect of Cross-View Contrastive Self-Supervision}. We also investigate the effectiveness of another core of \model's cross-view hypergraph-based contrastive learning. Specifically, we build a variant \emph{-CCL} by disabling the contrastive learning between the user-item interaction encoding and hypergraph dependency modeling. Another model variant \emph{-Hyper} only relies on the encoding of local collaborative relations to produce user and item representations. This variant does not capture the global user- and item-wise collaborative relationships through the hypergraph. \\\vspace{-0.12in}

\textbf{Results}. Clearly, \model\ always achieves the best performance as compared to competitive model variants, which further emphasizes the benefits of our hypergraph contrastive learning paradigm. To be specific, 1) our hypergraph structure learning is of great significance for the explicitly modeling of global property for user-item interaction patterns. It is in line with our assumption that our hypergraph neural network can alleviate the over-smoothing effect caused by local information aggregation. 2) The cross-view local-global contrastive learning paradigm indeed improves the performance of GNN-based collaborative filtering, with our self-supervised contrastive objectives. The incorporated intrinsic supervision labels reinforce the user-item interaction embedding space via the self-discrimination from local and global collaborative views.

\end{itemize}

\begin{table}[t]
    \caption{Ablation study on key components of \model.}
    \vspace{-0.15in}
    \centering
    \footnotesize
    \begin{tabular}{c|cc|cc|cc}
        \hline
        Data & \multicolumn{2}{c|}{Yelp} & \multicolumn{2}{c|}{MovieLens} & \multicolumn{2}{c}{Amazon}\\
        \hline
        Variants & Recall & NDCG & Recall & NDCG & Recall & NDCG\\
        \hline
        \hline
        \multicolumn{7}{c}{Top-20}\\
        \hline
        -LowR & 0.0587 & 0.0496 & 0.1937 & 0.2317 & 0.0319 & 0.0236\\
        -HHM & 0.0599 & 0.0505 & 0.1891 & 0.2319 & 0.0305 & 0.0230\\
        -Hyper & 0.0584 & 0.0490 & 0.1847 & 0.2233 & 0.0257 & 0.0188\\
        -CCL & 0.0566 & 0.0484 & 0.1984 & 0.2397 & 0.0282 & 0.0217\\
        \hline
        \emph{\model} & 0.0607 & 0.0510 & 0.2048 & 0.2467 & 0.0344 & 0.0258\\
        \hline
        \hline
        \multicolumn{7}{c}{Top-40}\\
        \hline
        -LowR & 0.0987 & 0.0634 & 0.2984 & 0.2622 & 0.0530 & 0.0309\\
        -HHM & 0.0992 & 0.0646 & 0.2914 & 0.2565 & 0.0505 & 0.0297\\
        -Hyper & 0.0982 & 0.0625 & 0.2780 & 0.2450 & 0.0419 & 0.0243\\
        -CCL & 0.0937 & 0.0619 & 0.3021 & 0.2650 & 0.0460 & 0.0276\\
        \hline
        \emph{\model} & 0.1007 & 0.0658 & 0.3081 & 0.2717 & 0.0561 & 0.0330\\
        \hline
    \end{tabular}
    \vspace{-0.15in}
    \label{tab:module_ablation}
\end{table}

\subsection{In Depth Analysis of \model's Benefits (RQ3)}
\subsubsection{\bf Robustness of \model\ in Alleviating Data Sparsity} 
\begin{figure}[t]
    \centering
    \subfigure[Yelp data]{
        \includegraphics[width=0.42\columnwidth]{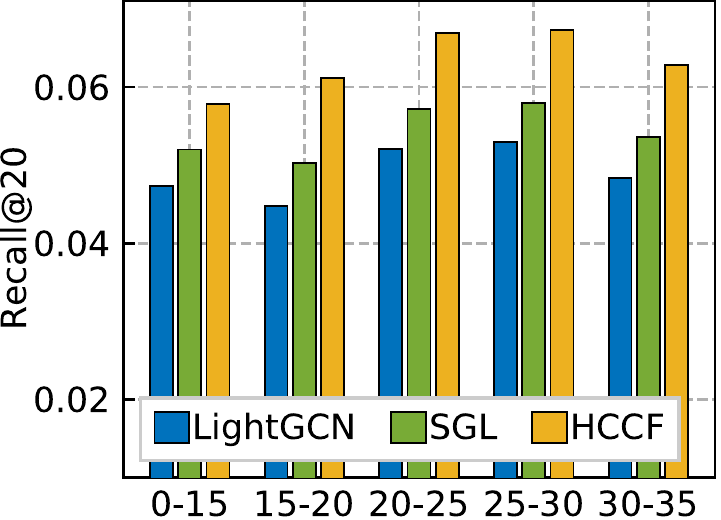}\quad
        \includegraphics[width=0.42\columnwidth]{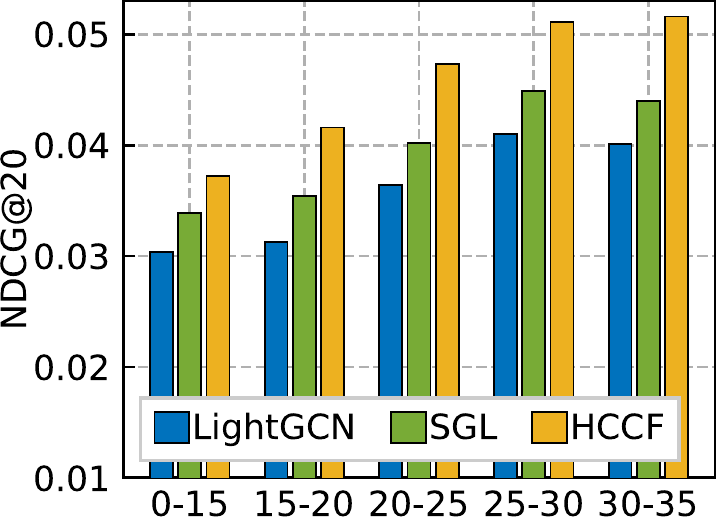}
        \vspace{-0.15in}
    }
    \subfigure[Amazon data]{
        \includegraphics[width=0.42\columnwidth]{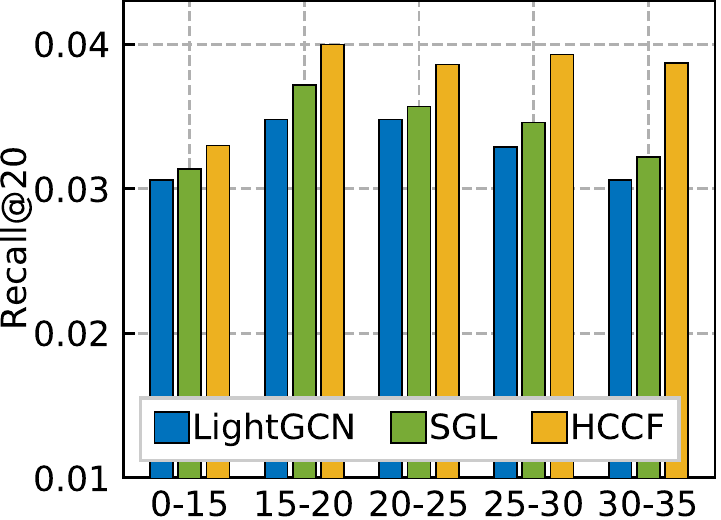}\quad 
        \includegraphics[width=0.42\columnwidth]{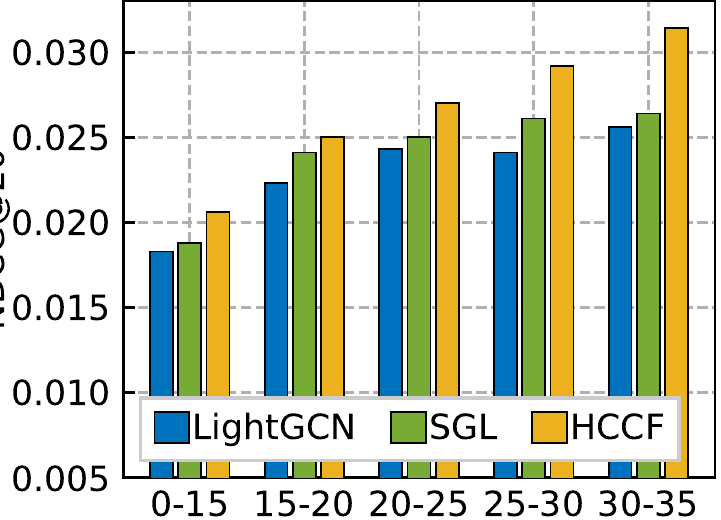}
    }
    \vspace{-0.2in}
    \caption{Performance \wrt\ interaction degrees.}
    \vspace{-0.2in}
    \label{fig:sparsity_amazon}
\end{figure}

To verify whether \model\ is robust to sparse issue which is ubiquitous in recommender systems, we partition users into different groups based on their interaction numbers (\eg, 20-25, 25-30). From the results shown in Figure~\ref{fig:sparsity_amazon}, our \model\ shows potentials in addressing the data sacristy issue. We ascribe this superiority to the \model's ability of cooperatively supervision between local collaborative relation encoding and global dependency learning. Furthermore, SGL is relatively unstable than \model\ across different sparsity degrees. It suggests that randomly dropping nodes or edges may discard important information of the original user-item interaction structures, making the training process unstable. The GCN-based method LightGCN may not learn quality representations for user preference by only relying on the sparse interaction data.

\subsubsection{\bf Effect of \model\ in Addressing Over-Smoothing}
With our designed cross-view hypergraph-guided contrastive learning component, user/item embeddings can be regularized to be far away based on their self-discrimination between the local-level and global-level collaborative relations. By doing so, the graph-based over-smoothing effect can be alleviated in our framework. To validate the effectiveness of our method in alleviating over-smoothing effect, in addition to the superior performance produced by our HCCF, we also calculate the Mean Average Distance (MAD)~\cite{chen2020measuring} over all node embedding pairs learned by the trained HCCF and two variants: i) -CCL (without the hypergraph-enhanced cross-view contrastive learning); ii) -Hyper (without the hypergraph neural network). The quantitative MAD metric measures the smoothness of a graph in terms of its node embeddings. The measurement results are shown in Table~\ref{tab:similarity}, where \textit{User} and \textit{Item} refer to the average similarity score between user nodes and item nodes, respectively. From the results, we can observe the more obvious over-smoothing phenomenon of the compared variants (-CCL) and (-Hyper) with smaller embedding distance scores. The above observations further justify the effectiveness of our hypergraph-enhanced contrastive learning in i) alleviating the over-smoothing issue in user representations refined with graph propagation; ii) empowering the model generalization ability of graph-based neural CF paradigm.

\begin{table}[h]
    \centering
    \vspace{-0.1in}
    \caption{Graph smoothness degrees (measured by MAD) with the encoded user/item embeddings by comparing with the variant -CCL (disabling the cross-view contrastive learning).}
    \vspace{-0.1in}
    \label{tab:similarity}
    \footnotesize
    \setlength{\tabcolsep}{1.1mm}
    \begin{tabular}{|c|c|c|c|c||c|c|c|c|c|}
        \hline
        Data & Type & -Hyper & -CCL & \emph{HCCF} & Data & Type & -Hyper & -CCL & \emph{HCCF}\\
        \hline
        \multirow{2}{*}{Yelp} & User & 0.9505 & 0.9106 & 0.9747 & \multirow{2}{*}{Amazon} & User & 0.7911 & 0.9106 & 0.9671\\
        \cline{2-5}
        \cline{7-10}
        & Item & 0.7673 & 0.9498 & 0.9671 & &  Item & 0.8570 & 0.7106 & 0.8573\\
        \hline
    \end{tabular}
    \vspace{-0.15in}
\end{table}


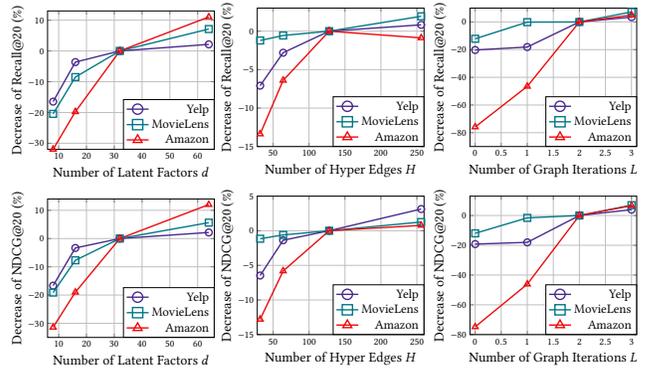
\begin{figure}
    \centering
    \begin{adjustbox}{max width=1.0\linewidth}
    \begin{tikzpicture}
\begin{axis}[
    xlabel={Number of Latent Factors $d$},
    ylabel={Decrease of Recall@20 (\%)},
    xmin=6,xmax=66,
    ymin=-32,ymax=13,
    legend columns=1,
    legend cell align=right,
    grid=both,
    every axis plot/.append style={ultra thick},
    every tick label/.append style={scale=1.3},
    label style={scale=1.8},
    legend style={
        nodes={scale=1.5, transform shape},
        legend image post style={scale=1.5},
        },
    legend style={at={(1,0)},anchor=south east},
    every axis plot post/.append style={
        every mark/.append style={scale=2}
    }
]
\addplot[color={rgb:red,133;green,76;blue,255}, mark=o, mark options={solid}]
table[x=para, y=yelp_hr] {latFactor.txt};
\addplot[color={rgb:red,0;green,157;blue,178}, mark=square, mark options={solid}]
table[x=para, y=ml10m_hr] {latFactor.txt};
\addplot[color={rgb:red,245;green,9;blue,11}, mark=triangle, mark options={solid}]
table[x=para, y=amazon_hr] {latFactor.txt};
\legend{\large Yelp, \large MovieLens, \large Amazon};
\end{axis}
\end{tikzpicture}

\begin{tikzpicture}
\begin{axis}[
    xlabel={Number of Hyper Edges $H$},
    ylabel={Decrease of Recall@20 (\%)},
    xmin=28,xmax=260,
    ymin=-15,ymax=3,
    legend columns=1,
    legend cell align=right,
    grid=both,
    every axis plot/.append style={ultra thick},
    every tick label/.append style={scale=1.3},
    label style={scale=1.8},
    legend style={
        nodes={scale=1.5, transform shape},
        legend image post style={scale=1.5},
        },
    legend style={at={(1,0)},anchor=south east},
    every axis plot post/.append style={
        every mark/.append style={scale=2}
    }
]
\addplot[color={rgb:red,133;green,76;blue,255}, mark=o, mark options={solid}]
table[x=para, y=yelp_hr] {hyperNum.txt};
\addplot[color={rgb:red,0;green,157;blue,178}, mark=square, mark options={solid}]
table[x=para, y=ml10m_hr] {hyperNum.txt};
\addplot[color={rgb:red,245;green,9;blue,11}, mark=triangle, mark options={solid}]
table[x=para, y=amazon_hr] {hyperNum.txt};
\legend{\large Yelp, \large MovieLens, \large Amazon};
\end{axis}
\end{tikzpicture}

\begin{tikzpicture}
\begin{axis}[
    xlabel={Number of Graph Iterations $L$},
    ylabel={Decrease of Recall@20 (\%)},
    xmin=-0.1,xmax=3.1,
    ymin=-90,ymax=10,
    legend columns=1,
    legend cell align=right,
    grid=both,
    every axis plot/.append style={ultra thick},
    every tick label/.append style={scale=1.3},
    label style={scale=1.8},
    legend style={
        nodes={scale=1.5, transform shape},
        legend image post style={scale=1.5},
        },
    legend style={at={(1,0)},anchor=south east},
    every axis plot post/.append style={
        every mark/.append style={scale=2}
    }
]
\addplot[color={rgb:red,133;green,76;blue,255}, mark=o, mark options={solid}]
table[x=para, y=yelp_hr] {gnnlayer.txt};
\addplot[color={rgb:red,0;green,157;blue,178}, mark=square, mark options={solid}]
table[x=para, y=ml10m_hr] {gnnlayer.txt};
\addplot[color={rgb:red,245;green,9;blue,11}, mark=triangle, mark options={solid}]
table[x=para, y=amazon_hr] {gnnlayer.txt};
\legend{\large Yelp, \large MovieLens, \large Amazon};
\end{axis}
\end{tikzpicture}
    \end{adjustbox}
    \begin{adjustbox}{max width=1.0\linewidth}
    \begin{tikzpicture}
\begin{axis}[
    xlabel={Number of Latent Factors $d$},
    ylabel={Decrease of NDCG@20 (\%)},
    xmin=6,xmax=66,
    ymin=-35,ymax=14,
    legend columns=1,
    legend cell align=right,
    grid=both,
    every axis plot/.append style={ultra thick},
    every tick label/.append style={scale=1.3},
    label style={scale=1.8},
    legend style={
        nodes={scale=1.5, transform shape},
        legend image post style={scale=1.5},
        },
    legend style={at={(1,0)},anchor=south east},
    every axis plot post/.append style={
        every mark/.append style={scale=2}
    }
]
\addplot[color={rgb:red,133;green,76;blue,255}, mark=o, mark options={solid}]
table[x=para, y=yelp_ndcg] {latFactor.txt};
\addplot[color={rgb:red,0;green,157;blue,178}, mark=square, mark options={solid}]
table[x=para, y=ml10m_ndcg] {latFactor.txt};
\addplot[color={rgb:red,245;green,9;blue,11}, mark=triangle, mark options={solid}]
table[x=para, y=amazon_ndcg] {latFactor.txt};
\legend{\large Yelp, \large MovieLens, \large Amazon};
\end{axis}
\end{tikzpicture}

\begin{tikzpicture}
\begin{axis}[
    xlabel={Number of Hyper Edges $H$},
    ylabel={Decrease of NDCG@20 (\%)},
    xmin=28,xmax=260,
    ymin=-15,ymax=5,
    legend columns=1,
    legend cell align=right,
    grid=both,
    every axis plot/.append style={ultra thick},
    every tick label/.append style={scale=1.3},
    label style={scale=1.8},
    legend style={
        nodes={scale=1.5, transform shape},
        legend image post style={scale=1.5},
        },
    legend style={at={(1,0)},anchor=south east},
    every axis plot post/.append style={
        every mark/.append style={scale=2}
    }
]
\addplot[color={rgb:red,133;green,76;blue,255}, mark=o, mark options={solid}]
table[x=para, y=yelp_ndcg] {hyperNum.txt};
\addplot[color={rgb:red,0;green,157;blue,178}, mark=square, mark options={solid}]
table[x=para, y=ml10m_ndcg] {hyperNum.txt};
\addplot[color={rgb:red,245;green,9;blue,11}, mark=triangle, mark options={solid}]
table[x=para, y=amazon_ndcg] {hyperNum.txt};
\legend{\large Yelp, \large MovieLens, \large Amazon};
\end{axis}
\end{tikzpicture}

\begin{tikzpicture}
\begin{axis}[
    xlabel={Number of Graph Iterations $L$},
    ylabel={Decrease of NDCG@20 (\%)},
    xmin=-0.1,xmax=3.1,
    ymin=-80,ymax=13,
    legend columns=1,
    legend cell align=right,
    grid=both,
    every axis plot/.append style={ultra thick},
    every tick label/.append style={scale=1.3},
    label style={scale=1.8},
    legend style={
        nodes={scale=1.5, transform shape},
        legend image post style={scale=1.5},
        },
    legend style={at={(1,0)},anchor=south east},
    every axis plot post/.append style={
        every mark/.append style={scale=2}
    }
]
\addplot[color={rgb:red,133;green,76;blue,255}, mark=o, mark options={solid}]
table[x=para, y=yelp_ndcg] {gnnlayer.txt};
\addplot[color={rgb:red,0;green,157;blue,178}, mark=square, mark options={solid}]
table[x=para, y=ml10m_ndcg] {gnnlayer.txt};
\addplot[color={rgb:red,245;green,9;blue,11}, mark=triangle, mark options={solid}]
table[x=para, y=amazon_ndcg] {gnnlayer.txt};
\legend{\large Yelp, \large MovieLens, \large Amazon};
\end{axis}
\end{tikzpicture}
    \end{adjustbox}
    \vspace{-0.25in}
    \caption{Hyperparameter study of the \model.}
    \vspace{-0.15in}
    \label{fig:hyperparam}
\end{figure}

\subsection{Hyperparameter Analysis (RQ4)}
\begin{figure}[t]
    \centering
    \includegraphics[width=0.46\columnwidth]{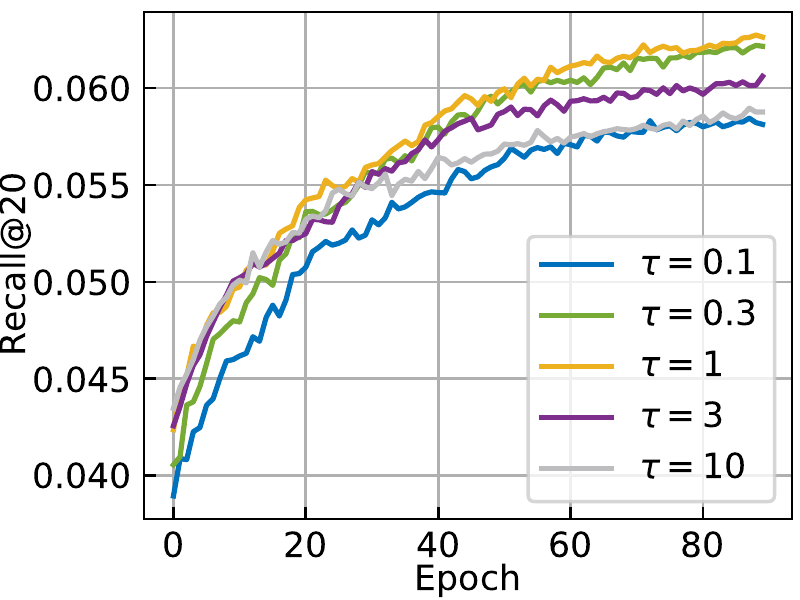}
    \includegraphics[width=0.46\columnwidth]{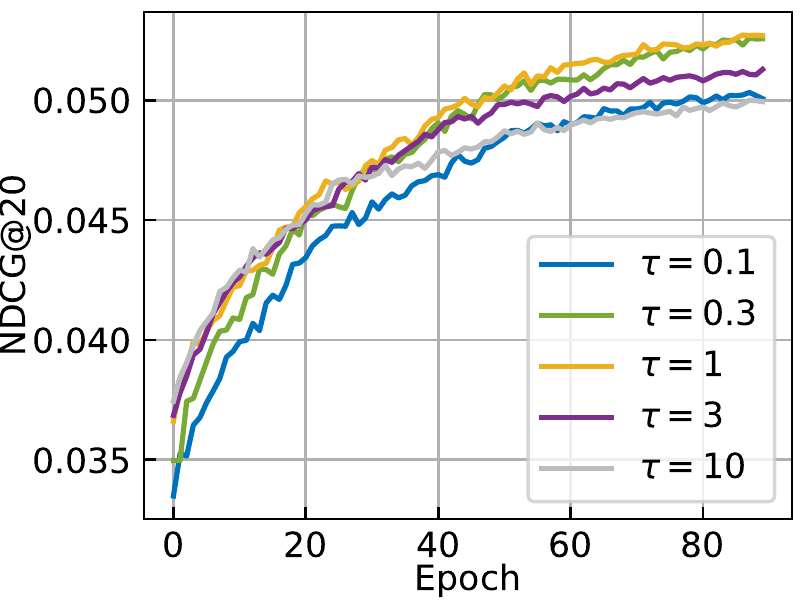}
    \vspace{-0.1in}
    \caption{Impact of $\tau$ in contrastive learning on Yelp data.}
    \label{fig:SSL_para}
    \vspace{-0.15in}
\end{figure}

In this section, we study the impact of several key hyperparameters (\eg, embedding dimensionality $d$, \# of hyperedge $H$, \# of graph message passing layers, temperature $\tau$) in our \model\ framework and report the evaluation results in Figure~\ref{fig:hyperparam} and Figure~\ref{fig:SSL_para}. 

\begin{itemize}[leftmargin=*]

\item (i) The best performance can be achieved with the hidden state dimensionality of 32 and the number of hyperedges of 128. In our hypergraph neural network, hyperedges serve as the intermediate connections acorss different users/items. Different hyperedge-specific cross-node structures may reflect different types of dependency semantics. Therefore, the recommendation performance degradation may stem from the overfitting issue with larger number of hyperedge-specific representation spaces.\\\vspace{-0.12in}

\item (ii) Two layers of graph encoder is sufficient to offer good performance, since our hypergraph-enhanced CF paradigm encourages the global collaborative relation modeling. Through the message passing across both adjacent and non-adjacent users under a hypergraph architecture, users with similar interaction patterns will be reinforced to achieve similar representations, so as to preserve the global collaborative context.\\\vspace{-0.12in}


\item (iii) In our framework of hypergraph-enhanced contrastive learning, the temperature parameter $\tau$ controls the strength of identifying hard negatives with the incorporated contrastive objective. From evaluation results in Figure~\ref{fig:SSL_para}, we can observe that the best performance can be achieved with $\tau=1.0$. Additionally, larger $\tau$ value ($\tau>1.0$) brings smaller gradient for learning hard negatives, which leads to the performance degradation.



\end{itemize}

\subsection{Case Study (RQ5)}
We qualitatively investigate the effects of our hypergraph-enhanced contrastive learning framework to i) capture the implicit global user dependency, and ii) alleviate the over-smoothing issue.

\subsubsection{\bf Global User Dependency}
\begin{figure}
    \centering
    \includegraphics[width=0.97\columnwidth]{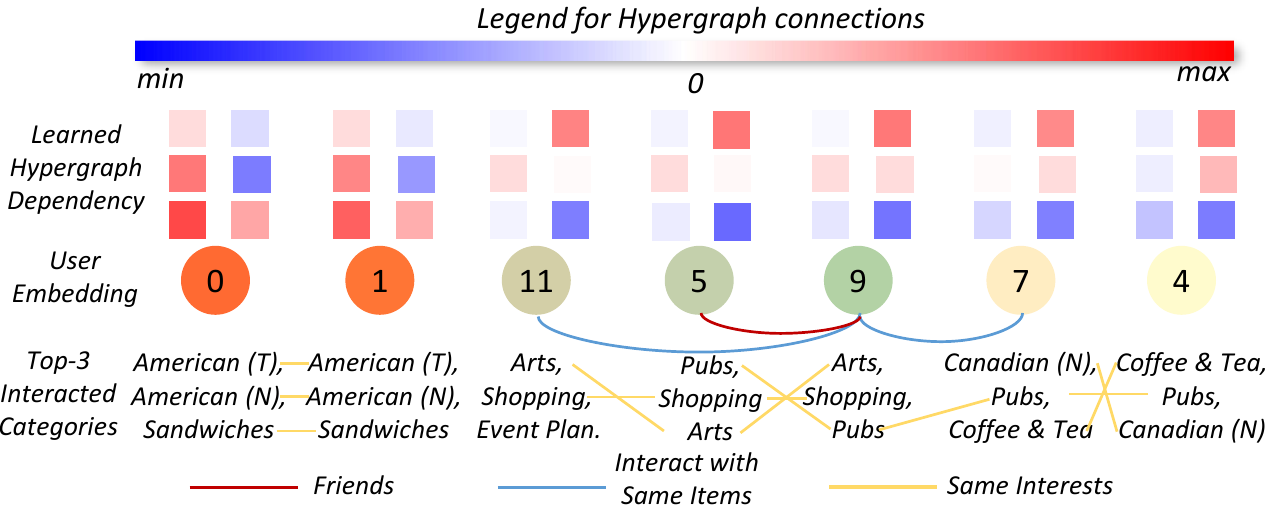}
    \vspace{-0.1in}
    \caption{Case study of capturing implicit global dependencies across users through our hypergraph learning.}
    \vspace{-0.12in}
    \label{fig:case_study}
\end{figure}

We project user embeddings into different colors based on the vector values. The user-hyperedge dependencies are presented with different colors in terms of the relevance scores. As shown in Figure~\ref{fig:case_study}, although the non-overlap interacted items between different users, our \model\ can distill their implicit dependency by generating similar embeddings (with similar node colors) only using user interaction data. For example, (1) the user pair with same interacted category flavors ($u_0$, $u_1$) and ($u_4$, $u_7$); (2) socially connected users ($u_5$, $u_9$). This provides an intuitive impression of \model's ability in exploring the implicit global user dependency for offering better recommendation performance.

\subsubsection{\bf Over-Smoothing Alleviation}
\begin{figure}[h]
    \centering
    \vspace{-0.1in}
    \includegraphics[width=0.87\columnwidth]{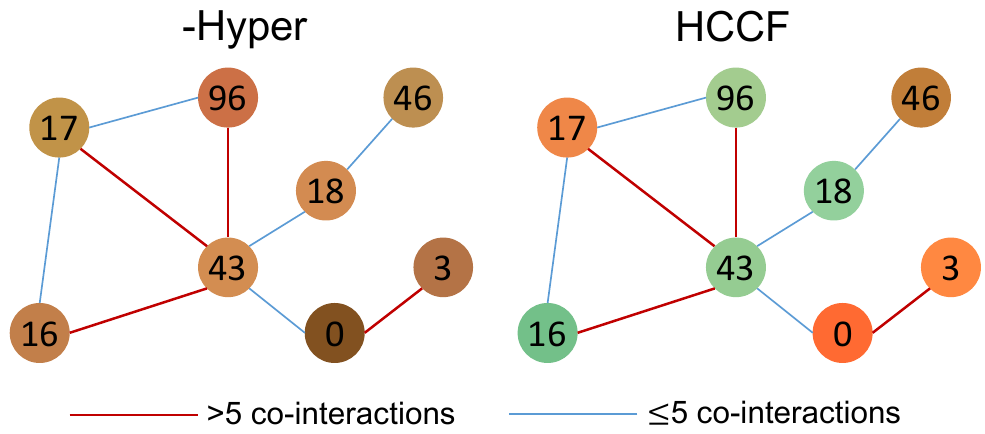}
    \vspace{-0.1in}
    \caption{Learned embeddings of users in a closely-connected sub-graph given by HCCF with and without our hypergraph learning compoent (model variant -Hyper).}
    \label{fig:rebuttal_case_study}
\end{figure}

we conduct case study to further investigate the ability of \model\ against the over-smoothing issue. We randomly pick a closely-connected sub-graph for users. The learned node embeddings and the co-interaction connections are shown in Figure~\ref{fig:rebuttal_case_study}. Here we also compare \model\ with the -Hyper variant. We can see that -Hyper assigns similar embeddings to all users in the sub-graph, as their embeddings are smoothed by each other. In contrast, \model\ is able to learn the subtle differences and divide the users into roughly two groups (colored with green and brown), even if two nodes are strongly bonded together (\eg, user 17 and user 43). By checking the detailed information about the sampled users, we found that the green users (16, 18, 43, 96) interact with much fewer items ($<50$ interactions) compared to the brown users ($\geq100$ interactions). Overall, \model\ is able to distinguish users with sparse and dense interactions.

\section{Conclusion}
\label{sec:conclusion}

This paper mainly focuses on enhancing the neural collaborative filtering with the hypergraph-guided self-supervised learning paradigm. We present \model\ which is featured by a hypergraph structure learning module and a cross-view hypergraph contrastive encoding schema. Our hypergraph contrastive CF framework learns better user representations by simultaneously characterizing both local and global collaborative relationships in a joint embedding space. Extensive experiments validate the superiority of \model\ towards competitive baselines. In future, we may explore the dynamic user dependency with a time-aware hypergraph embedding function, to inject temporal context into our CF architecture.

\section*{Acknowledgments}
This research is supported by the research grants from the Department of Computer Science \& Musketeers Foundation Institute of Data Science at the University of Hong Kong, the Natural Sciences \& Engineering Research Council (NSERC) of Canada.

\clearpage

\bibliographystyle{ACM-Reference-Format}
\balance
\bibliography{sigproc.bib}


\begin{thebibliography}{52}


\ifx \showCODEN    \undefined \def \showCODEN     #1{\unskip}     \fi
\ifx \showDOI      \undefined \def \showDOI       #1{#1}\fi
\ifx \showISBNx    \undefined \def \showISBNx     #1{\unskip}     \fi
\ifx \showISBNxiii \undefined \def \showISBNxiii  #1{\unskip}     \fi
\ifx \showISSN     \undefined \def \showISSN      #1{\unskip}     \fi
\ifx \showLCCN     \undefined \def \showLCCN      #1{\unskip}     \fi
\ifx \shownote     \undefined \def \shownote      #1{#1}          \fi
\ifx \showarticletitle \undefined \def \showarticletitle #1{#1}   \fi
\ifx \showURL      \undefined \def \showURL       {\relax}        \fi
\providecommand\bibfield[2]{#2}
\providecommand\bibinfo[2]{#2}
\providecommand\natexlab[1]{#1}
\providecommand\showeprint[2][]{arXiv:#2}

\bibitem[\protect\citeauthoryear{Berg, Kipf, and Welling}{Berg
  et~al\mbox{.}}{2017}]%
        {berg2017graph}
\bibfield{author}{\bibinfo{person}{Rianne van~den Berg},
  \bibinfo{person}{Thomas~N Kipf}, {and} \bibinfo{person}{Max Welling}.}
  \bibinfo{year}{2017}\natexlab{}.
\newblock \showarticletitle{Graph convolutional matrix completion}. In
  \bibinfo{booktitle}{\emph{KDD}}.
\newblock


\bibitem[\protect\citeauthoryear{Cao, Yang, and Ng}{Cao et~al\mbox{.}}{2021}]%
        {cao2021grammatical}
\bibfield{author}{\bibinfo{person}{Hannan Cao}, \bibinfo{person}{Wenmian Yang},
  {and} \bibinfo{person}{Hwee~Tou Ng}.} \bibinfo{year}{2021}\natexlab{}.
\newblock \showarticletitle{Grammatical Error Correction with Contrastive
  Learning in Low Error Density Domains}. In \bibinfo{booktitle}{\emph{EMNLP}}.
  \bibinfo{pages}{4867--4874}.
\newblock


\bibitem[\protect\citeauthoryear{Chen, Lin, Li, Li, Zhou, and Sun}{Chen
  et~al\mbox{.}}{2020b}]%
        {chen2020measuring}
\bibfield{author}{\bibinfo{person}{Deli Chen}, \bibinfo{person}{Yankai Lin},
  \bibinfo{person}{Wei Li}, \bibinfo{person}{Peng Li}, \bibinfo{person}{Jie
  Zhou}, {and} \bibinfo{person}{Xu Sun}.} \bibinfo{year}{2020}\natexlab{b}.
\newblock \showarticletitle{Measuring and relieving the over-smoothing problem
  for graph neural networks from the topological view}. In
  \bibinfo{booktitle}{\emph{AAAI}}, Vol.~\bibinfo{volume}{34}.
  \bibinfo{pages}{3438--3445}.
\newblock


\bibitem[\protect\citeauthoryear{Chen, Wu, Hong, Zhang, and Wang}{Chen
  et~al\mbox{.}}{2020c}]%
        {chen2020revisiting}
\bibfield{author}{\bibinfo{person}{Lei Chen}, \bibinfo{person}{Le Wu},
  \bibinfo{person}{Richang Hong}, \bibinfo{person}{Kun Zhang}, {and}
  \bibinfo{person}{Meng Wang}.} \bibinfo{year}{2020}\natexlab{c}.
\newblock \showarticletitle{Revisiting Graph Based Collaborative Filtering: A
  Linear Residual Graph Convolutional Network Approach}. In
  \bibinfo{booktitle}{\emph{AAAI}}, Vol.~\bibinfo{volume}{34}.
  \bibinfo{pages}{27--34}.
\newblock


\bibitem[\protect\citeauthoryear{Chen, Kornblith, Norouzi, and Hinton}{Chen
  et~al\mbox{.}}{2020a}]%
        {chen2020simple}
\bibfield{author}{\bibinfo{person}{Ting Chen}, \bibinfo{person}{Simon
  Kornblith}, \bibinfo{person}{Mohammad Norouzi}, {and}
  \bibinfo{person}{Geoffrey Hinton}.} \bibinfo{year}{2020}\natexlab{a}.
\newblock \showarticletitle{A simple framework for contrastive learning of
  visual representations}. In \bibinfo{booktitle}{\emph{ICML}}. PMLR,
  \bibinfo{pages}{1597--1607}.
\newblock


\bibitem[\protect\citeauthoryear{Chen, Xiong, Zhang, Xia, et~al\mbox{.}}{Chen
  et~al\mbox{.}}{2020d}]%
        {chen2020neural}
\bibfield{author}{\bibinfo{person}{Xu Chen}, \bibinfo{person}{Kun Xiong},
  \bibinfo{person}{Yongfeng Zhang}, \bibinfo{person}{Long Xia},
  {et~al\mbox{.}}} \bibinfo{year}{2020}\natexlab{d}.
\newblock \showarticletitle{Neural Feature-aware Recommendation with Signed
  Hypergraph Convolutional Network}.
\newblock \bibinfo{journal}{\emph{TOIS}} \bibinfo{volume}{39},
  \bibinfo{number}{1} (\bibinfo{year}{2020}), \bibinfo{pages}{1--22}.
\newblock


\bibitem[\protect\citeauthoryear{Dong, Yu, Wu, Sun, Yuan, and Zhang}{Dong
  et~al\mbox{.}}{2017}]%
        {dong2017hybrid}
\bibfield{author}{\bibinfo{person}{Xin Dong}, \bibinfo{person}{Lei Yu},
  \bibinfo{person}{Zhonghuo Wu}, \bibinfo{person}{Yuxia Sun},
  \bibinfo{person}{Lingfeng Yuan}, {and} \bibinfo{person}{Fangxi Zhang}.}
  \bibinfo{year}{2017}\natexlab{}.
\newblock \showarticletitle{A hybrid collaborative filtering model with deep
  structure for recommender systems}. In \bibinfo{booktitle}{\emph{AAAI}},
  Vol.~\bibinfo{volume}{31}.
\newblock


\bibitem[\protect\citeauthoryear{Feng, You, Zhang, Ji, and Gao}{Feng
  et~al\mbox{.}}{2019}]%
        {feng2019hypergraph}
\bibfield{author}{\bibinfo{person}{Yifan Feng}, \bibinfo{person}{Haoxuan You},
  \bibinfo{person}{Zizhao Zhang}, \bibinfo{person}{Rongrong Ji}, {and}
  \bibinfo{person}{Yue Gao}.} \bibinfo{year}{2019}\natexlab{}.
\newblock \showarticletitle{Hypergraph neural networks}. In
  \bibinfo{booktitle}{\emph{AAAI}}, Vol.~\bibinfo{volume}{33}.
  \bibinfo{pages}{3558--3565}.
\newblock


\bibitem[\protect\citeauthoryear{Gao, Zhang, Lin, Zhao, Du, and Zou}{Gao
  et~al\mbox{.}}{2020}]%
        {gao2020hypergraph}
\bibfield{author}{\bibinfo{person}{Yue Gao}, \bibinfo{person}{Zizhao Zhang},
  \bibinfo{person}{Haojie Lin}, \bibinfo{person}{Xibin Zhao},
  \bibinfo{person}{Shaoyi Du}, {and} \bibinfo{person}{Changqing Zou}.}
  \bibinfo{year}{2020}\natexlab{}.
\newblock \showarticletitle{Hypergraph learning: Methods and practices}.
\newblock \bibinfo{journal}{\emph{TPAMI}} (\bibinfo{year}{2020}).
\newblock


\bibitem[\protect\citeauthoryear{Guo, Zhuang, Qin, Zhu, Xie, Xiong, and He}{Guo
  et~al\mbox{.}}{2020}]%
        {guo2020survey}
\bibfield{author}{\bibinfo{person}{Qingyu Guo}, \bibinfo{person}{Fuzhen
  Zhuang}, \bibinfo{person}{Chuan Qin}, \bibinfo{person}{Hengshu Zhu},
  \bibinfo{person}{Xing Xie}, \bibinfo{person}{Hui Xiong}, {and}
  \bibinfo{person}{Qing He}.} \bibinfo{year}{2020}\natexlab{}.
\newblock \showarticletitle{A survey on knowledge graph-based recommender
  systems}.
\newblock \bibinfo{journal}{\emph{TKDE}} (\bibinfo{year}{2020}).
\newblock


\bibitem[\protect\citeauthoryear{He, Chen, Wang, Jameel, Yu, and Xu}{He
  et~al\mbox{.}}{2021}]%
        {he2021click}
\bibfield{author}{\bibinfo{person}{Li He}, \bibinfo{person}{Hongxu Chen},
  \bibinfo{person}{Dingxian Wang}, \bibinfo{person}{Shoaib Jameel},
  \bibinfo{person}{Philip Yu}, {and} \bibinfo{person}{Guandong Xu}.}
  \bibinfo{year}{2021}\natexlab{}.
\newblock \showarticletitle{Click-Through Rate Prediction with Multi-Modal
  Hypergraphs}. In \bibinfo{booktitle}{\emph{CIKM}}. \bibinfo{pages}{690--699}.
\newblock


\bibitem[\protect\citeauthoryear{He, Deng, Wang, Li, Zhang, and Wang}{He
  et~al\mbox{.}}{2020}]%
        {he2020lightgcn}
\bibfield{author}{\bibinfo{person}{Xiangnan He}, \bibinfo{person}{Kuan Deng},
  \bibinfo{person}{Xiang Wang}, \bibinfo{person}{Yan Li},
  \bibinfo{person}{Yongdong Zhang}, {and} \bibinfo{person}{Meng Wang}.}
  \bibinfo{year}{2020}\natexlab{}.
\newblock \showarticletitle{Lightgcn: Simplifying and powering graph
  convolution network for recommendation}. In
  \bibinfo{booktitle}{\emph{SIGIR}}. \bibinfo{pages}{639--648}.
\newblock


\bibitem[\protect\citeauthoryear{He, Liao, Zhang, Nie, Hu, and Chua}{He
  et~al\mbox{.}}{2017}]%
        {he2017neural}
\bibfield{author}{\bibinfo{person}{Xiangnan He}, \bibinfo{person}{Lizi Liao},
  \bibinfo{person}{Hanwang Zhang}, \bibinfo{person}{Liqiang Nie},
  \bibinfo{person}{Xia Hu}, {and} \bibinfo{person}{Tat-Seng Chua}.}
  \bibinfo{year}{2017}\natexlab{}.
\newblock \showarticletitle{Neural collaborative filtering}. In
  \bibinfo{booktitle}{\emph{WWW}}. \bibinfo{pages}{173--182}.
\newblock


\bibitem[\protect\citeauthoryear{He, Zhang, Kan, and Chua}{He
  et~al\mbox{.}}{2016}]%
        {he2016fast}
\bibfield{author}{\bibinfo{person}{Xiangnan He}, \bibinfo{person}{Hanwang
  Zhang}, \bibinfo{person}{Min-Yen Kan}, {and} \bibinfo{person}{Tat-Seng
  Chua}.} \bibinfo{year}{2016}\natexlab{}.
\newblock \showarticletitle{Fast matrix factorization for online recommendation
  with implicit feedback}. In \bibinfo{booktitle}{\emph{SIGIR}}.
  \bibinfo{pages}{549--558}.
\newblock


\bibitem[\protect\citeauthoryear{Huang}{Huang}{2021}]%
        {huang2021recent}
\bibfield{author}{\bibinfo{person}{Chao Huang}.}
  \bibinfo{year}{2021}\natexlab{}.
\newblock \showarticletitle{Recent Advances in Heterogeneous Relation Learning
  for Recommendation}.
\newblock \bibinfo{journal}{\emph{arXiv preprint arXiv:2110.03455}}
  (\bibinfo{year}{2021}).
\newblock


\bibitem[\protect\citeauthoryear{Huang, Chen, Xia, Xu, Dai, Chen, Bo, Zhao, and
  Huang}{Huang et~al\mbox{.}}{2021}]%
        {huang2021graph}
\bibfield{author}{\bibinfo{person}{Chao Huang}, \bibinfo{person}{Jiahui Chen},
  \bibinfo{person}{Lianghao Xia}, \bibinfo{person}{Yong Xu},
  \bibinfo{person}{Peng Dai}, \bibinfo{person}{Yanqing Chen},
  \bibinfo{person}{Liefeng Bo}, \bibinfo{person}{Jiashu Zhao}, {and}
  \bibinfo{person}{Jimmy~Xiangji Huang}.} \bibinfo{year}{2021}\natexlab{}.
\newblock \showarticletitle{Graph-enhanced multi-task learning of multi-level
  transition dynamics for session-based recommendation}. In
  \bibinfo{booktitle}{\emph{AAAI}}.
\newblock


\bibitem[\protect\citeauthoryear{Huang, Wu, Zhang, Zhang, Zhao, Yin, and
  Chawla}{Huang et~al\mbox{.}}{2019}]%
        {2019online}
\bibfield{author}{\bibinfo{person}{Chao Huang}, \bibinfo{person}{Xian Wu},
  \bibinfo{person}{Xuchao Zhang}, \bibinfo{person}{Chuxu Zhang},
  \bibinfo{person}{Jiashu Zhao}, \bibinfo{person}{Dawei Yin}, {and}
  \bibinfo{person}{Nitesh~V Chawla}.} \bibinfo{year}{2019}\natexlab{}.
\newblock \showarticletitle{Online purchase prediction via multi-scale modeling
  of behavior dynamics}. In \bibinfo{booktitle}{\emph{KDD}}.
  \bibinfo{pages}{2613--2622}.
\newblock


\bibitem[\protect\citeauthoryear{Ji, Feng, Ji, Zhao, Tang, and Gao}{Ji
  et~al\mbox{.}}{2020}]%
        {ji2020dual}
\bibfield{author}{\bibinfo{person}{Shuyi Ji}, \bibinfo{person}{Yifan Feng},
  \bibinfo{person}{Rongrong Ji}, \bibinfo{person}{Xibin Zhao},
  \bibinfo{person}{Wanwan Tang}, {and} \bibinfo{person}{Yue Gao}.}
  \bibinfo{year}{2020}\natexlab{}.
\newblock \showarticletitle{Dual channel hypergraph collaborative filtering}.
  In \bibinfo{booktitle}{\emph{KDD}}. \bibinfo{pages}{2020--2029}.
\newblock


\bibitem[\protect\citeauthoryear{Jiang, Wei, Feng, Cao, and Gao}{Jiang
  et~al\mbox{.}}{2019}]%
        {jiang2019dynamic}
\bibfield{author}{\bibinfo{person}{Jianwen Jiang}, \bibinfo{person}{Yuxuan
  Wei}, \bibinfo{person}{Yifan Feng}, \bibinfo{person}{Jingxuan Cao}, {and}
  \bibinfo{person}{Yue Gao}.} \bibinfo{year}{2019}\natexlab{}.
\newblock \showarticletitle{Dynamic Hypergraph Neural Networks.}. In
  \bibinfo{booktitle}{\emph{IJCAI}}. \bibinfo{pages}{2635--2641}.
\newblock


\bibitem[\protect\citeauthoryear{Khosla, Teterwak, Wang, Sarna, Tian, Isola,
  Maschinot, Liu, and Krishnan}{Khosla et~al\mbox{.}}{2020}]%
        {khosla2020supervised}
\bibfield{author}{\bibinfo{person}{Prannay Khosla}, \bibinfo{person}{Piotr
  Teterwak}, \bibinfo{person}{Chen Wang}, \bibinfo{person}{Aaron Sarna},
  \bibinfo{person}{Yonglong Tian}, \bibinfo{person}{Phillip Isola},
  \bibinfo{person}{Aaron Maschinot}, \bibinfo{person}{Ce Liu}, {and}
  \bibinfo{person}{Dilip Krishnan}.} \bibinfo{year}{2020}\natexlab{}.
\newblock \showarticletitle{Supervised contrastive learning}.
\newblock \bibinfo{journal}{\emph{NIPS}}  \bibinfo{volume}{33}
  (\bibinfo{year}{2020}), \bibinfo{pages}{18661--18673}.
\newblock


\bibitem[\protect\citeauthoryear{Koren, Bell, et~al\mbox{.}}{Koren
  et~al\mbox{.}}{2009}]%
        {koren2009matrix}
\bibfield{author}{\bibinfo{person}{Yehuda Koren}, \bibinfo{person}{Robert
  Bell}, {et~al\mbox{.}}} \bibinfo{year}{2009}\natexlab{}.
\newblock \showarticletitle{Matrix factorization techniques for recommender
  systems}.
\newblock \bibinfo{journal}{\emph{Computer}} \bibinfo{number}{8}
  (\bibinfo{year}{2009}), \bibinfo{pages}{30--37}.
\newblock


\bibitem[\protect\citeauthoryear{Long, Huang, Xu, Xu, Dai, Xia, and Bo}{Long
  et~al\mbox{.}}{2021}]%
        {long2021social}
\bibfield{author}{\bibinfo{person}{Xiaoling Long}, \bibinfo{person}{Chao
  Huang}, \bibinfo{person}{Yong Xu}, \bibinfo{person}{Huance Xu},
  \bibinfo{person}{Peng Dai}, \bibinfo{person}{Lianghao Xia}, {and}
  \bibinfo{person}{Liefeng Bo}.} \bibinfo{year}{2021}\natexlab{}.
\newblock \showarticletitle{Social Recommendation with Self-Supervised
  Metagraph Informax Network}. In \bibinfo{booktitle}{\emph{CIKM}}.
  \bibinfo{pages}{1160--1169}.
\newblock


\bibitem[\protect\citeauthoryear{Ma, Cui, Kuang, Wang, and Zhu}{Ma
  et~al\mbox{.}}{2019a}]%
        {ma2019disentangled}
\bibfield{author}{\bibinfo{person}{Jianxin Ma}, \bibinfo{person}{Peng Cui},
  \bibinfo{person}{Kun Kuang}, \bibinfo{person}{Xin Wang}, {and}
  \bibinfo{person}{Wenwu Zhu}.} \bibinfo{year}{2019}\natexlab{a}.
\newblock \showarticletitle{Disentangled graph convolutional networks}. In
  \bibinfo{booktitle}{\emph{ICML}}. PMLR, \bibinfo{pages}{4212--4221}.
\newblock


\bibitem[\protect\citeauthoryear{Ma, Zhou, Cui, Yang, and Zhu}{Ma
  et~al\mbox{.}}{2019b}]%
        {ma2019learning}
\bibfield{author}{\bibinfo{person}{Jianxin Ma}, \bibinfo{person}{Chang Zhou},
  \bibinfo{person}{Peng Cui}, \bibinfo{person}{Hongxia Yang}, {and}
  \bibinfo{person}{Wenwu Zhu}.} \bibinfo{year}{2019}\natexlab{b}.
\newblock \showarticletitle{Learning disentangled representations for
  recommendation}. In \bibinfo{booktitle}{\emph{NIPS}}.
  \bibinfo{pages}{5711--5722}.
\newblock


\bibitem[\protect\citeauthoryear{Min, Wenkel, and Wolf}{Min
  et~al\mbox{.}}{2020}]%
        {min2020scattering}
\bibfield{author}{\bibinfo{person}{Yimeng Min}, \bibinfo{person}{Frederik
  Wenkel}, {and} \bibinfo{person}{Guy Wolf}.} \bibinfo{year}{2020}\natexlab{}.
\newblock \showarticletitle{Scattering gcn: Overcoming oversmoothness in graph
  convolutional networks}.
\newblock  (\bibinfo{year}{2020}).
\newblock


\bibitem[\protect\citeauthoryear{Mnih and Salakhutdinov}{Mnih and
  Salakhutdinov}{2008}]%
        {mnih2008probabilistic}
\bibfield{author}{\bibinfo{person}{Andriy Mnih} {and} \bibinfo{person}{Russ~R
  Salakhutdinov}.} \bibinfo{year}{2008}\natexlab{}.
\newblock \showarticletitle{Probabilistic matrix factorization}. In
  \bibinfo{booktitle}{\emph{NIPS}}. \bibinfo{pages}{1257--1264}.
\newblock


\bibitem[\protect\citeauthoryear{Oord, Li, and Vinyals}{Oord
  et~al\mbox{.}}{2018}]%
        {oord2018representation}
\bibfield{author}{\bibinfo{person}{Aaron van~den Oord}, \bibinfo{person}{Yazhe
  Li}, {and} \bibinfo{person}{Oriol Vinyals}.} \bibinfo{year}{2018}\natexlab{}.
\newblock \showarticletitle{Representation learning with contrastive predictive
  coding}.
\newblock \bibinfo{journal}{\emph{arXiv preprint arXiv:1807.03748}}
  (\bibinfo{year}{2018}).
\newblock


\bibitem[\protect\citeauthoryear{Peng, Wang, Desrosiers, and Pedersoli}{Peng
  et~al\mbox{.}}{2021}]%
        {peng2021self}
\bibfield{author}{\bibinfo{person}{Jizong Peng}, \bibinfo{person}{Ping Wang},
  \bibinfo{person}{Christian Desrosiers}, {and} \bibinfo{person}{Marco
  Pedersoli}.} \bibinfo{year}{2021}\natexlab{}.
\newblock \showarticletitle{Self-paced contrastive learning for semi-supervised
  medical image segmentation with meta-labels}.
\newblock \bibinfo{journal}{\emph{NIPS}}  \bibinfo{volume}{34}
  (\bibinfo{year}{2021}).
\newblock


\bibitem[\protect\citeauthoryear{Qiu, Chen, Dong, Zhang, Yang, Ding, Wang, and
  Tang}{Qiu et~al\mbox{.}}{2020}]%
        {qiu2020gcc}
\bibfield{author}{\bibinfo{person}{Jiezhong Qiu}, \bibinfo{person}{Qibin Chen},
  \bibinfo{person}{Yuxiao Dong}, \bibinfo{person}{Jing Zhang},
  \bibinfo{person}{Hongxia Yang}, \bibinfo{person}{Ming Ding},
  \bibinfo{person}{Kuansan Wang}, {and} \bibinfo{person}{Jie Tang}.}
  \bibinfo{year}{2020}\natexlab{}.
\newblock \showarticletitle{Gcc: Graph contrastive coding for graph neural
  network pre-training}. In \bibinfo{booktitle}{\emph{KDD}}.
  \bibinfo{pages}{1150--1160}.
\newblock


\bibitem[\protect\citeauthoryear{Sedhain, Menon, Sanner, and Xie}{Sedhain
  et~al\mbox{.}}{2015}]%
        {sedhain2015autorec}
\bibfield{author}{\bibinfo{person}{Suvash Sedhain},
  \bibinfo{person}{Aditya~Krishna Menon}, \bibinfo{person}{Scott Sanner}, {and}
  \bibinfo{person}{Lexing Xie}.} \bibinfo{year}{2015}\natexlab{}.
\newblock \showarticletitle{Autorec: Autoencoders meet collaborative
  filtering}. In \bibinfo{booktitle}{\emph{WWW}}. \bibinfo{pages}{111--112}.
\newblock


\bibitem[\protect\citeauthoryear{Stojanovski, Krojer, Peskov, and
  Fraser}{Stojanovski et~al\mbox{.}}{2020}]%
        {stojanovski2020contracat}
\bibfield{author}{\bibinfo{person}{Dario Stojanovski}, \bibinfo{person}{Benno
  Krojer}, \bibinfo{person}{Denis Peskov}, {and} \bibinfo{person}{Alexander
  Fraser}.} \bibinfo{year}{2020}\natexlab{}.
\newblock \showarticletitle{ContraCAT: Contrastive coreference analytical
  templates for machine translation}. In \bibinfo{booktitle}{\emph{COLING}}.
  \bibinfo{pages}{4732--4749}.
\newblock


\bibitem[\protect\citeauthoryear{Sun, Cheng, Zuberi, P{\'e}rez, and
  Volkovs}{Sun et~al\mbox{.}}{2021}]%
        {sun2021hgcf}
\bibfield{author}{\bibinfo{person}{Jianing Sun}, \bibinfo{person}{Zhaoyue
  Cheng}, \bibinfo{person}{Saba Zuberi}, \bibinfo{person}{Felipe P{\'e}rez},
  {and} \bibinfo{person}{Maksims Volkovs}.} \bibinfo{year}{2021}\natexlab{}.
\newblock \showarticletitle{HGCF: Hyperbolic Graph Convolution Networks for
  Collaborative Filtering}. In \bibinfo{booktitle}{\emph{WWW}}.
  \bibinfo{pages}{593--601}.
\newblock


\bibitem[\protect\citeauthoryear{Wang, Ding, Hong, Liu, and Caverlee}{Wang
  et~al\mbox{.}}{2020a}]%
        {wang2020next}
\bibfield{author}{\bibinfo{person}{Jianling Wang}, \bibinfo{person}{Kaize
  Ding}, \bibinfo{person}{Liangjie Hong}, \bibinfo{person}{Huan Liu}, {and}
  \bibinfo{person}{James Caverlee}.} \bibinfo{year}{2020}\natexlab{a}.
\newblock \showarticletitle{Next-item recommendation with sequential
  hypergraphs}. In \bibinfo{booktitle}{\emph{SIGIR}}.
  \bibinfo{pages}{1101--1110}.
\newblock


\bibitem[\protect\citeauthoryear{Wang, Chen, Zhu, Shen, and Zhang}{Wang
  et~al\mbox{.}}{2019a}]%
        {wang2019unified}
\bibfield{author}{\bibinfo{person}{Pengfei Wang}, \bibinfo{person}{Hanxiong
  Chen}, \bibinfo{person}{Yadong Zhu}, \bibinfo{person}{Huawei Shen}, {and}
  \bibinfo{person}{Yongfeng Zhang}.} \bibinfo{year}{2019}\natexlab{a}.
\newblock \showarticletitle{Unified collaborative filtering over graph
  embeddings}. In \bibinfo{booktitle}{\emph{SIGIR}}. \bibinfo{pages}{155--164}.
\newblock


\bibitem[\protect\citeauthoryear{Wang, He, Wang, Feng, and Chua}{Wang
  et~al\mbox{.}}{2019b}]%
        {wang2019neural}
\bibfield{author}{\bibinfo{person}{Xiang Wang}, \bibinfo{person}{Xiangnan He},
  \bibinfo{person}{Meng Wang}, \bibinfo{person}{Fuli Feng}, {and}
  \bibinfo{person}{Tat-Seng Chua}.} \bibinfo{year}{2019}\natexlab{b}.
\newblock \showarticletitle{Neural Graph Collaborative Filtering}. In
  \bibinfo{booktitle}{\emph{SIGIR}}.
\newblock


\bibitem[\protect\citeauthoryear{Wang, Jin, Zhang, He, Xu, and Chua}{Wang
  et~al\mbox{.}}{2020b}]%
        {wang2020disentangled}
\bibfield{author}{\bibinfo{person}{Xiang Wang}, \bibinfo{person}{Hongye Jin},
  \bibinfo{person}{An Zhang}, \bibinfo{person}{Xiangnan He},
  \bibinfo{person}{Tong Xu}, {and} \bibinfo{person}{Tat-Seng Chua}.}
  \bibinfo{year}{2020}\natexlab{b}.
\newblock \showarticletitle{Disentangled graph collaborative filtering}. In
  \bibinfo{booktitle}{\emph{SIGIR}}. \bibinfo{pages}{1001--1010}.
\newblock


\bibitem[\protect\citeauthoryear{Wei, Huang, Xia, Xu, Zhao, and Yin}{Wei
  et~al\mbox{.}}{2022}]%
        {wei2022contrastive}
\bibfield{author}{\bibinfo{person}{Wei Wei}, \bibinfo{person}{Chao Huang},
  \bibinfo{person}{Lianghao Xia}, \bibinfo{person}{Yong Xu},
  \bibinfo{person}{Jiashu Zhao}, {and} \bibinfo{person}{Dawei Yin}.}
  \bibinfo{year}{2022}\natexlab{}.
\newblock \showarticletitle{Contrastive Meta Learning with Behavior
  Multiplicity for Recommendation}. In \bibinfo{booktitle}{\emph{WSDM}}.
  \bibinfo{pages}{1120--1128}.
\newblock


\bibitem[\protect\citeauthoryear{Wu, Wang, Feng, He, Chen, Lian, and Xie}{Wu
  et~al\mbox{.}}{2021}]%
        {wu2021self}
\bibfield{author}{\bibinfo{person}{Jiancan Wu}, \bibinfo{person}{Xiang Wang},
  \bibinfo{person}{Fuli Feng}, \bibinfo{person}{Xiangnan He},
  \bibinfo{person}{Liang Chen}, \bibinfo{person}{Jianxun Lian}, {and}
  \bibinfo{person}{Xing Xie}.} \bibinfo{year}{2021}\natexlab{}.
\newblock \showarticletitle{Self-supervised graph learning for recommendation}.
  In \bibinfo{booktitle}{\emph{SIGIR}}. \bibinfo{pages}{726--735}.
\newblock


\bibitem[\protect\citeauthoryear{Wu, DuBois, Zheng, and Ester}{Wu
  et~al\mbox{.}}{2016}]%
        {wu2016collaborative}
\bibfield{author}{\bibinfo{person}{Yao Wu}, \bibinfo{person}{Christopher
  DuBois}, \bibinfo{person}{Alice~X Zheng}, {and} \bibinfo{person}{Martin
  Ester}.} \bibinfo{year}{2016}\natexlab{}.
\newblock \showarticletitle{Collaborative denoising auto-encoders for top-n
  recommender systems}. In \bibinfo{booktitle}{\emph{WSDM}}. ACM,
  \bibinfo{pages}{153--162}.
\newblock


\bibitem[\protect\citeauthoryear{Xia, Huang, Xu, Dai, Zhang, Yang, Pei, and
  Bo}{Xia et~al\mbox{.}}{2021a}]%
        {xia2021knowledge}
\bibfield{author}{\bibinfo{person}{Lianghao Xia}, \bibinfo{person}{Chao Huang},
  \bibinfo{person}{Yong Xu}, \bibinfo{person}{Peng Dai}, \bibinfo{person}{Xiyue
  Zhang}, \bibinfo{person}{Hongsheng Yang}, \bibinfo{person}{Jian Pei}, {and}
  \bibinfo{person}{Liefeng Bo}.} \bibinfo{year}{2021}\natexlab{a}.
\newblock \showarticletitle{Knowledge-enhanced hierarchical graph transformer
  network for multi-behavior recommendation}. In
  \bibinfo{booktitle}{\emph{AAAI}}, Vol.~\bibinfo{volume}{35}.
  \bibinfo{pages}{4486--4493}.
\newblock


\bibitem[\protect\citeauthoryear{Xia, Xu, Huang, Dai, and Bo}{Xia
  et~al\mbox{.}}{2021b}]%
        {xia2021graph}
\bibfield{author}{\bibinfo{person}{Lianghao Xia}, \bibinfo{person}{Yong Xu},
  \bibinfo{person}{Chao Huang}, \bibinfo{person}{Peng Dai}, {and}
  \bibinfo{person}{Liefeng Bo}.} \bibinfo{year}{2021}\natexlab{b}.
\newblock \showarticletitle{Graph meta network for multi-behavior
  recommendation}. In \bibinfo{booktitle}{\emph{SIGIR}}.
  \bibinfo{pages}{757--766}.
\newblock


\bibitem[\protect\citeauthoryear{Xue, Dai, Zhang, Huang, and Chen}{Xue
  et~al\mbox{.}}{2017}]%
        {xue2017deep}
\bibfield{author}{\bibinfo{person}{Hong-Jian Xue}, \bibinfo{person}{Xinyu Dai},
  \bibinfo{person}{Jianbing Zhang}, \bibinfo{person}{Shujian Huang}, {and}
  \bibinfo{person}{Jiajun Chen}.} \bibinfo{year}{2017}\natexlab{}.
\newblock \showarticletitle{Deep Matrix Factorization Models for Recommender
  Systems.}. In \bibinfo{booktitle}{\emph{IJCAI}}. \bibinfo{pages}{3203--3209}.
\newblock


\bibitem[\protect\citeauthoryear{Yao, Yi, Cheng, et~al\mbox{.}}{Yao
  et~al\mbox{.}}{2021}]%
        {yao2021self}
\bibfield{author}{\bibinfo{person}{Tiansheng Yao}, \bibinfo{person}{Xinyang
  Yi}, \bibinfo{person}{Derek~Zhiyuan Cheng}, {et~al\mbox{.}}}
  \bibinfo{year}{2021}\natexlab{}.
\newblock \showarticletitle{Self-supervised Learning for Large-scale Item
  Recommendations}. In \bibinfo{booktitle}{\emph{CIKM}}.
  \bibinfo{pages}{4321--4330}.
\newblock


\bibitem[\protect\citeauthoryear{Ying, He, Chen, Eksombatchai, Hamilton, and
  Leskovec}{Ying et~al\mbox{.}}{2018}]%
        {ying2018graph}
\bibfield{author}{\bibinfo{person}{Rex Ying}, \bibinfo{person}{Ruining He},
  \bibinfo{person}{Kaifeng Chen}, \bibinfo{person}{Pong Eksombatchai},
  \bibinfo{person}{William~L Hamilton}, {and} \bibinfo{person}{Jure Leskovec}.}
  \bibinfo{year}{2018}\natexlab{}.
\newblock \showarticletitle{Graph convolutional neural networks for web-scale
  recommender systems}. In \bibinfo{booktitle}{\emph{KDD}}.
  \bibinfo{pages}{974--983}.
\newblock


\bibitem[\protect\citeauthoryear{Yu, Yin, Li, Wang, Hung, and Zhang}{Yu
  et~al\mbox{.}}{2021}]%
        {yu2021self}
\bibfield{author}{\bibinfo{person}{Junliang Yu}, \bibinfo{person}{Hongzhi Yin},
  \bibinfo{person}{Jundong Li}, \bibinfo{person}{Qinyong Wang},
  \bibinfo{person}{Nguyen Quoc~Viet Hung}, {and} \bibinfo{person}{Xiangliang
  Zhang}.} \bibinfo{year}{2021}\natexlab{}.
\newblock \showarticletitle{Self-Supervised Multi-Channel Hypergraph
  Convolutional Network for Social Recommendation}. In
  \bibinfo{booktitle}{\emph{WWW}}. \bibinfo{pages}{413--424}.
\newblock


\bibitem[\protect\citeauthoryear{Yuhao, Huang, Xia, and Li}{Yuhao
  et~al\mbox{.}}{2022}]%
        {kgcontrastive2022}
\bibfield{author}{\bibinfo{person}{Yang Yuhao}, \bibinfo{person}{Chao Huang},
  \bibinfo{person}{Lianghao Xia}, {and} \bibinfo{person}{Chenliang Li}.}
  \bibinfo{year}{2022}\natexlab{}.
\newblock \showarticletitle{Knowledge graph contrastive learning for
  recommendation}. In \bibinfo{booktitle}{\emph{SIGIR}}.
\newblock


\bibitem[\protect\citeauthoryear{Zhang, Shi, Zhao, and King}{Zhang
  et~al\mbox{.}}{2019}]%
        {zhang2019star}
\bibfield{author}{\bibinfo{person}{Jiani Zhang}, \bibinfo{person}{Xingjian
  Shi}, \bibinfo{person}{Shenglin Zhao}, {and} \bibinfo{person}{Irwin King}.}
  \bibinfo{year}{2019}\natexlab{}.
\newblock \showarticletitle{Star-gcn: Stacked and reconstructed graph
  convolutional networks for recommender systems}. In
  \bibinfo{booktitle}{\emph{IJCAI}}.
\newblock


\bibitem[\protect\citeauthoryear{Zhao, Wang, Shi, Hu, Song, and Ye}{Zhao
  et~al\mbox{.}}{2021}]%
        {zhao2021heterogeneous}
\bibfield{author}{\bibinfo{person}{Jianan Zhao}, \bibinfo{person}{Xiao Wang},
  \bibinfo{person}{Chuan Shi}, \bibinfo{person}{Binbin Hu},
  \bibinfo{person}{Guojie Song}, {and} \bibinfo{person}{Yanfang Ye}.}
  \bibinfo{year}{2021}\natexlab{}.
\newblock \showarticletitle{Heterogeneous Graph Structure Learning for Graph
  Neural Networks}. In \bibinfo{booktitle}{\emph{AAAI}}.
\newblock


\bibitem[\protect\citeauthoryear{Zheng, Li, Lu, Zhang, and Yu}{Zheng
  et~al\mbox{.}}{2019}]%
        {zheng2019deep}
\bibfield{author}{\bibinfo{person}{Lei Zheng}, \bibinfo{person}{Chaozhuo Li},
  \bibinfo{person}{Chun-Ta Lu}, \bibinfo{person}{Jiawei Zhang}, {and}
  \bibinfo{person}{Philip~S Yu}.} \bibinfo{year}{2019}\natexlab{}.
\newblock \showarticletitle{Deep Distribution Network: Addressing the Data
  Sparsity Issue for Top-N Recommendation}. In
  \bibinfo{booktitle}{\emph{SIGIR}}. \bibinfo{pages}{1081--1084}.
\newblock


\bibitem[\protect\citeauthoryear{Zhou, Huang, Li, Zha, Chen, and Hu}{Zhou
  et~al\mbox{.}}{2020}]%
        {zhou2020towards}
\bibfield{author}{\bibinfo{person}{Kaixiong Zhou}, \bibinfo{person}{Xiao
  Huang}, \bibinfo{person}{Yuening Li}, \bibinfo{person}{Daochen Zha},
  \bibinfo{person}{Rui Chen}, {and} \bibinfo{person}{Xia Hu}.}
  \bibinfo{year}{2020}\natexlab{}.
\newblock \showarticletitle{Towards deeper graph neural networks with
  differentiable group normalization}. In \bibinfo{booktitle}{\emph{NIPS}}.
\newblock


\bibitem[\protect\citeauthoryear{Zhu, Xu, Yu, Liu, Wu, and Wang}{Zhu
  et~al\mbox{.}}{2021}]%
        {zhu2021graph}
\bibfield{author}{\bibinfo{person}{Yanqiao Zhu}, \bibinfo{person}{Yichen Xu},
  \bibinfo{person}{Feng Yu}, \bibinfo{person}{Qiang Liu}, \bibinfo{person}{Shu
  Wu}, {and} \bibinfo{person}{Liang Wang}.} \bibinfo{year}{2021}\natexlab{}.
\newblock \showarticletitle{Graph contrastive learning with adaptive
  augmentation}. In \bibinfo{booktitle}{\emph{WWW}}.
  \bibinfo{pages}{2069--2080}.
\newblock


\bibitem[\protect\citeauthoryear{Zou, Xia, Gu, Zhao, Liu, Huang, and Yin}{Zou
  et~al\mbox{.}}{2020}]%
        {zou2020neural}
\bibfield{author}{\bibinfo{person}{Lixin Zou}, \bibinfo{person}{Long Xia},
  \bibinfo{person}{Yulong Gu}, \bibinfo{person}{Xiangyu Zhao},
  \bibinfo{person}{Weidong Liu}, \bibinfo{person}{Jimmy~Xiangji Huang}, {and}
  \bibinfo{person}{Dawei Yin}.} \bibinfo{year}{2020}\natexlab{}.
\newblock \showarticletitle{Neural interactive collaborative filtering}. In
  \bibinfo{booktitle}{\emph{SIGIR}}. \bibinfo{pages}{749--758}.
\newblock


\end{thebibliography}

\end{document}